\documentclass[twocolumn,epjc3]{svjour3}  
\smartqed  
\RequirePackage{graphicx}
\RequirePackage[numbers,sort&compress]{natbib}
\RequirePackage{amssymb}
\RequirePackage{amsmath}
\usepackage{mathbbol}
\usepackage{bm}
\usepackage[figure,longend]{algorithm2e}
\usepackage{color}
\usepackage{dsfont}
\usepackage{braket}

\newcommand{\dd}{\mathrm{d}}

\newcommand{\p}{\partial}

\newcommand{\transp}{^{\text{T}}}

\newcommand{\inv}{^{-1}}

\newcommand{\dt}{\delta t}

\newcommand{\angstrom}{\mbox{\normalfont\AA}}
\newcommand{\hartree}{\mbox{\normalfont $E_\mathrm{h}$}}
\newcommand{\elec}{\mbox{\normalfont $e$}}
\newcommand{\um}[1]{\ensuremath{\mathrm{\,#1}}} 

\renewcommand{\vec}{\bm}

\DeclareSymbolFontAlphabet{\mathbb}{bbold}
\DeclareSymbolFontAlphabet{\mathbbm}{AMSb}

\journalname{Eur. Phys. J. B}
\begin{document}

\title{Mass-Zero constrained dynamics and statistics for the shell model in magnetic field
}


\author{D. D. Girardier\thanksref{addr1}
        \and
        A. Coretti\thanksref{addr2,addr1} 
        \and
        G. Ciccotti\thanksref{addr3,addr4,addr5} 
        \and
        S. Bonella\thanksref{addr1,e1}
}

\thankstext{e1}{e-mail: sara.bonella@epfl.ch}


\institute{Centre Europ\'een de Calcul Atomique et Mol\'eculaire (CECAM), Ecole Polytechnique F\'ed\'erale de Lausanne, 1015 Lausanne, Switzerland \label{addr1}
           \and
           Department of Mathematical Sciences, Politecnico di Torino, Corso Duca degli Abruzzi 24, I-10129 Torino, Italy \label{addr2}
           \and
           Institute for Applied Computing ``Mauro Picone'' (IAC), CNR Via dei Taurini 19, 00185 Rome, Italy \label{addr3}
           \and
           Universit\`a di Roma La Sapienza, Ple. A. Moro 5, 00185 Roma, Italy \label{addr4}
           \and
           School of Physics, University College of Dublin UCD-Belfield, Dublin 4, Ireland \label{addr5}
}

\date{Received: date / Accepted: date}

\maketitle

\begin{abstract}
In several domains of physics, including \emph{first principle} simulations and classical models for polarizable systems, the minimization of an energy function with respect to a set of auxiliary variables must be performed to define the dynamics of physical degrees of freedom. In this paper, we discuss a recent algorithm proposed to efficiently and rigorously simulate this type of systems: the Mass-Zero (MaZe) Constrained Dynamics. In MaZe the minimum condition is imposed as a constraint on the auxiliary variables treated as degrees of freedom of zero inertia driven by the physical system. The method is formulated in the Lagrangian framework, enabling the properties of the approach to emerge naturally from a fully consistent dynamical and statistical viewpoint. We begin by presenting MaZe for typical minimization problems where the imposed constraints are holonomic and summarizing its key formal properties, notably the exact Born-Oppenheimer dynamics followed by the physical variables and the exact sampling of the corresponding physical probability density. We then generalize the approach to the case of conditions on the auxiliary variables that linearly involve their velocities. Such conditions occur, for example, when describing systems in external magnetic field and they require to adapt MaZe to integrate semiholonomic constraints. The new development is presented in the second part of this paper and illustrated via a proof-of-principle calculation of the charge transport properties of a simple classical polarizable model of NaCl. 

\end{abstract}

\section{Introduction}
\label{intro}
In this paper, we discuss, focusing on recent developments, the Mass-Zero (MaZe) constrained dynamics and further extend it to a simple but interesting model of classical polarizable systems in constant external magnetic field. MaZe is a general simulation approach to study the motion of a set of physical degrees of freedom (dofs) whose evolution depends on parameters subject to given conditions. The method considers an extended system in which the parameters appear as (auxiliary) dynamical variables together with the original dofs and the conditions are interpreted as constraints. The coupled evolution equations for the overall constrained system are then conveniently obtained in the Lagrangian formalism. From these, the original parametric dynamics for the physical dofs is rigorously recovered by taking the limit of zero mass for the auxiliary variables. In practical implementations, use of the SHAKE algorithm enables symplectic and efficient numerical integration of the extended dynamical system.

The method of mass-zero constraints was originally introduced in the early 1980's to study the rotational-translational coupling in diatomic molecules~\cite{ryckaert:1981}. Recently, it has undergone a new set of developments when adiabatic systems were identified as an important area where MaZe dynamics can provide an original formal approach and an effective integration algorithm~\cite{coretti:2018b,bonella:2020,coretti:2020a}. In adiabatic systems, the substantial timescale separation of the motion of two sets of interacting degrees of freedom justifies adopting the Born-Oppenheimer approximation for the evolution. The timescale gap is typically due to the disparate masses of the two sets and full adiabatic separation is achieved in the limit of zero mass for the fast dofs. In the context of nuclear and electronic motion, for which the Born-Oppenheimer approximation was originally introduced, adiabaticity also requires the existence of a finite gap between the HOMO and LUMO electronic states. In the full adiabatic regime, evolution equations, typically of classical form, are given for the slow degrees of freedom. The forces on the slow variables, however, depend parametrically on the values of the fast variables. These values are obtained, for each configuration of the slow degrees of freedom along the trajectory, enforcing the condition that the interaction potential (a function of both sets) is at a minimum with respect to the fast dofs. In Molecular Dynamics (MD), a typical example of adiabatic dynamics is the evolution of ionic (slow) and electronic (fast) degrees of freedom in \emph{first principle} calculations based on Kohn-Sham or orbital-free Density Functional Theory. Another important example is given by classical models of polarization in which electrons do not appear directly, but the dynamical system is extended to include sets of classical auxiliary variables of null mass mimicking different polarization effects.

Current methods adopted for the MD simulation of such systems combine standard propagation schemes for the slow variables --- that we shall indicate as the ions for simplicity --- with algorithms to find, or approximate, the minimum of the potential with respect to the fast dofs at each ionic configuration. Depending on the specific system, additional conditions, such as orthonormality or sum rules, may be imposed on the fast dofs, affecting the minimum search. Traditional schemes for finding the minimum include iterative methods, notably the conjugate-gradient approach~\cite{aguado:2003a,jahn:2004} adopted in Born-Oppenheimer MD. In Car-Parrinello MD, on the other hand, an extended system in which the auxiliary dofs are treated as dynamical variables with a small mass is defined~\cite{sprik:1988,wilson:1993}. In this scheme, the minimum condition is approximately tracked, with a precision that improves with smaller mass for the auxiliary dofs~\cite{payne:1986,pastore:1991,payne:1992,marx:2012-book}, via the dynamics itself thus avoiding iterations. More recently, alternative \textit{ad hoc} dynamics for the auxiliary variables like the so-called always stable predictor-corrector approach~\cite{kolafa:2004,genzer:2004} have also been employed. 
All these methods, however, suffer from practical or conceptual limitations. Conjugate-gradient minimization is guaranteed to converge only in the case of a quadratic function to be minimized, and, for the general minimization problems typically associated with realistic condensed-phase models, can be unstable~\cite{pacaud:2018} or expensive~\cite{pounds:2009} to fully converge. Incomplete convergence of the iterative minimization in Born-Oppenheimer MD, also known as the self-consistent-field optimization, has been shown to cause energy transfer between the slow and fast dofs~\cite{remler:1990,pulay:2004}, leading to energy drift in the ionic propagation and hindering access to long simulation timescales. Energy transfer, and the consequent violation of the adiabatic separation in the system, affects also Car-Parrinello propagation~\cite{aguado:2003a} due to the finite ratio of the masses associated to the fast and slow dofs. This pathology is often mitigated via thermostats that, however, may affect or, in the worst cases, compromise correct statistical sampling. Furthermore, the Car-Parrinello algorithm requires a very small timestep to integrate accurately the dynamics of the fast variables. The always stable predictor-corrector scheme is only approximately time reversible~\cite{kolafa:2004,genzer:2004} leading again to energy drifts that are usually quenched via a Berendsen thermostat (thus raising questions on the ensemble sampled by the dynamics), and it contains system dependent parameters that can only be determined by trial and error. An alternative scheme, that combines an extended Lagrangian approach with efficient self-consistent minimization, was recently proposed to address these shortcomings in the so-called extended Lagrangian Born-Oppenheimer approach~\cite{niklasson:2006,niklasson:2020}. Implementations of this approach, however, are nontrivial. The method relies on the introduction of auxiliary dynamical variables and of an approximate energy function in order to obtain stable and time-reversible evolution. The cost of computing the approximate energy increases substantially with the system size, leading so far to applications to relatively small systems or simple Lagrangians. Furthermore, like Car-Parrinello MD, this approach requires a timestep smaller than the one necessary for purely ionic evolution and, while the method is in principle fully symplectic and time-reversible, in practice problems arising from the accumulation of errors lead to implementations that do not rigorously preserve these properties~\cite{niklasson:2021}. 

The MaZe approach avoids many of the difficulties described above. Adopting the framework of constrained MD~\cite{ciccotti:1986}, in combination with the SHAKE algorithm~\cite{ryckaert:1977}, MaZe enables to derive and numerically solve classical evolution equations for an extended system that rigorously enforces exact adiabatic propagation. Exploiting and adapting the formalism of Lagrange multipliers, the method can incorporate easily additional constraints that depend only on the auxiliary variables, such as electroneutrality in classical polarizable models or orthonormality in \emph{first principles} calculations. The MaZe dynamical system is solved via a fully symplectic, time-reversible algorithm that guarantees stability of the evolution on the same timescale and with the same timestep size of standard MD for the physical dofs. The algorithm prevents, by construction, propagation of the error when imposing the constraints. Furthermore, the approach avoids standard self-consistent cycles for the minimization along the dynamics and uses iterations only to solve the equation of constraints, a process that usually has fast convergence, in particular in nonlinear problems, enabling to reach essentially the numerical precision limit at an affordable cost~\cite{coretti:2018b,bonella:2020,coretti:2020a}. Importantly, rigorous sampling of the target probability for the physical dofs (which coincides  with the one usually associated to Born-Oppenheimer dynamics) is also guaranteed~\cite{bonella:2020} ensuring that not only the adiabatic evolution, but also the exact statistical mechanics of the system is obtained. In the context of adiabatic systems, MaZe has been used to simulate simple classical polarizable systems~\cite{coretti:2018b}, and more recently for  state-of-the-art classical modelling of electrode charges in electrochemical systems~\cite{coretti:2020a} (the method has also been implemented in MetalWalls~\cite{marin-lafleche:2020}, a high-performance community software in this area). The generalization to first principle MD based on orbital-free density functional theory was also derived and tested with very good results~\cite{bonella:2020}. 

In the following, we first discuss the key aspects of MaZe summarizing recent work on the dynamical and statistical properties of the approach. For simplicity, the formalism will be presented using as reference application the adiabatic evolution of classical polarizable models. This choice is motivated also by the new MaZe development presented in the second part of the paper: the generalization of the approach for adiabatic propagation of classical polarizable models in external magnetic field. This generalization is prompted by the problem of simulating ionic charge transport in systems subject to a magnetic field, with particular focus on the ionic Hall effect~\cite{meton:1976,newman:1977,stuhrmann:2002}, an area that has attracted considerable interest to investigate the properties of superionic conductors~\cite{dieterich:1985,funke:2013} and, more recently, to enhance the capacitance of batteries~\cite{tang:2019}. Including the magnetic field in  classical polarizable models requires some nontrivial adaptation of the MaZe approach. Firstly, due to the Lorentz force, the condition of null force on the fast degrees of freedom involves both the coordinates and the velocities of these particles, thus leading to a set of \emph{nonholonomic} constraints associated to the MaZe dynamics. These constraints are, however, linear in the velocities, enabling to adapt the approach via a relatively standard generalization of the Lagrangian equations of motion for the system. Secondly, again due to the dependence of the constraint on the velocities, the standard SHAKE algorithm --- not the idea at the basis of the approach --- cannot be directly applied. While some methods for the numerical integration of systems subject to nonholonomic constraints exist~\cite{kutteh:1999,kneller:2017}, they are not as consolidated as SHAKE and often rely on nonsymplectic algorithms. In this work, we then propose an appropriate integration algorithm and test its properties. 
\\ 
The paper is organized as follows. In Section~\ref{sec:MaZeReview} the derivation of the MaZe dynamical system and the proof of exact sampling of the Born-Oppenheimer probability density for the slow degrees of freedom are summarized. The presentation will be self-contained, also with the support of~\ref{app:MaZeStatMech}, but we limit the proofs to their key steps, referring to previous work for more details. Our focus, in fact, will be to illustrate the most interesting formal and practical aspects of the approach. In Section~\ref{sec:MagneticMaZeShell} we then introduce the generalization of the MaZe framework to the case of classical polarizable simulations in constant magnetic field, together with the new algorithm for the solution of the nonholonomic constrained dynamics. We consider, for convenience, the simplest model of classical polarization, i.e. the shell model~\cite{dick:1958}. Section~\ref{sec:Results} reports an illustrative calculation investigating the combined effect of polarization and of the magnetic field on ionic charge transport properties in liquid NaCl. 

\section{MaZe dynamics and statistical mechanics}
As mentioned above, to simplify the presentation and tackle an interesting physical case, we illustrate the MaZe approach via its application to classical polarizable models. These models are commonly employed to simulate systems of theoretical and technological interest such as devices for electrochemical energy storage~\cite{simon:2008,armand:2008,beguin:2014} in which large sizes and long timescale prevent direct calculation of polarization effects via the quantum treatment of the electronic density in \emph{first principles} MD. In this area then, polarization effects are described by constructing empirical potentials that include sets of auxiliary variables that mimic changes in the electronic density. 
Because they represent electronic properties, these auxiliary variables are assumed to adapt instantaneously to the ionic configuration in the spirit of the Born-Oppenheimer approximation and are assigned a null mass. An early example of this type of models is the shell model~\cite{dick:1958,jacucci:1974,jacucci:1976}, which accounts for dipole polarization. Potentials that take into account interactions due to quadrupoles~\cite{wilson:1996a} and changes in the ions size and shape~\cite{wilson:1996b,rowley:1998} have also been introduced. More recently, models for capacitors have been proposed that include the mutual polarization of the elements combining a multipole description of the electrolyte with the so-called fluctuating charge model~\cite{sprik:1988,scalfi:2020} for the electrodes.

To be more specific, let us indicate with $\vec{R} \in \mathbbm{R}^{3N}$ the Cartesian coordinates of the $N$ physical dofs in the system (ions), and with $s \in \mathbbm{R}^M$ the $M$ adiabatically separated auxiliary variables. Depending on the specific polarization model, the $s$ variables may represent positions (as in the shell model) or different types of degrees of freedom (e.g. dipoles or quadrupoles, or charges) and their physical dimensions and number vary accordingly. The two sets of variables interact via the  potential $V(\vec{R}, s)$. 
The adiabatic dynamics of the system is obtained by imposing that the auxiliary variables satisfy, for all values of $\vec{R}$ along the trajectory, the condition
\begin{equation}
\label{eq:ConstraintConditions}
\frac{\partial V(\vec{R}, s)}{\partial s_\alpha} = 0 \qquad \alpha = 1,\dots,M
\end{equation}
Additional conditions, such as the charge neutrality constraint for classical models of electrodes, may be associated to the auxiliary dofs. These conditions are typically expressed as 
\begin{equation}
\label{eq:AdditionalConstraints}
f_\iota(s) =  0 \quad (\iota = 1, \dots, C)
\end{equation}
where $C$ is the number of the additional constraints that we shall assume to be, as it is often the case, functions of the $s$ variables alone. Eq.~\eqref{eq:AdditionalConstraints} implies that not all variations of the $s$ are independent, and this must be accounted for in the search of the minimum of the potential. For example, in Born-Oppenheimer dynamics, the conjugate-gradient minimization is conducted via a constrained search while in Car-Parrinello schemes, the additional constraints are added in the evolution equations derived from the Lagrangian. In the following, we indicate with $\tilde{s}$ the values of the auxiliary variables satisfying both the minimum condition on the potential and the additional constraints, if they are present. Due to the dependence of the potential on $\vec{R}$, $\tilde{s}=\tilde{s}(\vec{R})$. In the adiabatic regime, then, the evolution of the physical variables is given by
\begin{equation}
\label{eq:BODynamicsPolar}
m_i\ddot{\vec{R}}_i = -\nabla_{\vec{R}_i}V(\vec{R}, s)\big|_{s=\tilde{s}} 
\end{equation}
where $m_i$ is the mass of ion $i$. As discussed in the Introduction, current approaches for the numerical solution of the equation above have limitations that justify the development of alternative schemes. Starting from the next subsection, the MaZe approach is presented.

\label{sec:MaZeReview}
\subsection{MaZe dynamical system}
\label{sec:MaZeDynamics}
The key idea of the mass-zero constrained dynamics is to construct a Lagrangian that includes the $s$ as dynamical variables and to interpret Eq.~\eqref{eq:ConstraintConditions} as a set of holonomic constraints. In this section, MaZe is presented for the general case in which additional conditions must be satisfied by the fast dofs. In this case, the system is further extended to include also the Lagrange multipliers associated to these conditions as auxiliary variables. To set the stage, let us introduce the auxiliary function
\begin{equation}\label{eq:AuxPot}
W(\vec{R}, s, \nu)\equiv V(\vec{R}, s)+\sum_{\iota = 1}^C\nu_\iota f_\iota(s)
\end{equation}
where $\nu = \{ \nu_1,...,\nu_C\}$ are Lagrange multipliers associated to the additional constraints. The solution for $s$ and $\nu$ satisfying Eqs.~\eqref{eq:ConstraintConditions} and ~\eqref{eq:AdditionalConstraints} is then given by the stationary point $(\hat{s},\hat{\nu})$ of $W(\vec{R}, s, \nu)$~\cite{lanczos:1986-book,allaire:2007-book}. This leads to the $M+C$ conditions
\begin{equation}
\label{eq:SigmaNkPiusNu}
\begin{aligned}
\sigma_\alpha(\vec{R}, s, \nu) &= \frac{\partial W(\vec{R}, s, \nu)}{\partial s_\alpha} = 0  \,\,\,\, (\alpha=1,\dots,M) \\
\sigma_{M+\iota}(\vec{R}, s, \nu) &= \frac{\partial W(\vec{R}, s, \nu)}{\partial \nu_\iota} = 0  \,\,\,\, (\iota=1,\dots,C) .
\end{aligned}
\end{equation}
Note that the last set of equations above represents in fact the additional conditions Eq.~\eqref{eq:AdditionalConstraints}, now obtained as a result of an optimization problem in the space that includes the Lagrange multipliers $\nu$ as additional variables. In the absence of additional conditions, on the other hand, $W(\vec{R}, s, \nu)\equiv V(\vec{R}, s)$ and the $M$ surviving conditions above reduce to Eq.~\eqref{eq:ConstraintConditions}. At this stage, a finite fictitious ``mass'' is assigned to both sets of auxiliary variables.\footnote{Note that, depending on the physical dimensions of the auxiliary variables, the fictitious mass $\mu$ could have different units than those of a mass.} Indicating the fictitious mass for the $s$ and $\nu$ variables as $\mu_s$ and $\mu_{\nu}$, respectively, the Lagrangian for the extended system is defined as
\begin{eqnarray}
\label{eq:LNkPiusNu}
L(\vec{R}, \dot{\vec{R}}, s, \dot{s}, \nu, \dot{\nu}) &=& \frac{1}{2}\sum_{i=1}^{N}m_i\dot{\vec{R}}^{2}_{i} + \frac{1}{2}\sum_{\alpha=1}^M\mu_s \dot{s}^2_\alpha \nonumber \\
&+& \frac{1}{2}\sum_{\iota=1}^C\mu_{\nu} \dot{\nu}^2_{\iota} - W(\vec{R}, s, \nu),
\end{eqnarray}
From this, the constrained evolution equations are obtained as
\begin{equation}
\label{eq:MZ_nozerolimit}
\begin{aligned}
m_i\ddot{\vec{R}}_i &= -\nabla_{\vec{R}_i}W(\vec{R}, s, \nu)  - \sum_{\beta = 1}^{M+C} \lambda_{\beta} \nabla_{\vec{R}_i}\sigma_{\beta}(\vec{R}, s, \nu)\\
\mu_s\ddot{s}_\alpha &= -\frac{\p W(\vec{R}, s, \nu)}{\p s_{\alpha}} - \sum_{\beta = 1}^{M+C} \lambda_{\beta} \frac{\p \sigma_{\beta}(\vec{R}, s, \nu)}{\p s_{\alpha}}\\
\mu_{\nu}\ddot{\nu}_{\iota} &= -\frac{\p W(\vec{R}, s, \nu)}{\p \nu_{\iota}} -\sum_{\beta =1}^{M+C}\lambda_{\beta} \frac{\partial \sigma_{\beta}(\vec{R}, s, \nu)}{\partial \nu_\iota}
\end{aligned}
\end{equation}
The equations above can be simplified by observing that, in the first line, $\nabla_{\vec{R}_i}W(\vec{R}, s, \nu) =\nabla_{\vec{R}_i}V(\vec{R}, s)$, and that the forces acting on the auxiliary variables are null because they coincide with the constraints. Thus, dividing both sides of the equations for the auxiliary variables by their masses,
\begin{equation}
\label{eq:MZ_stepone}
\begin{aligned}
m_i\ddot{\vec{R}}_i &= -\nabla_{\vec{R}_i}V(\vec{R}, s)  - \sum_{\beta = 1}^{M+C} \lambda_{\beta} \nabla_{\vec{R}_i}\sigma_{\beta}(\vec{R}, s, \nu)\\
\ddot{s}_\alpha &= - \sum_{\beta = 1}^{M+C} \frac{\lambda_{\beta}}{\mu_s} \frac{\p \sigma_{\beta}(\vec{R}, s, \nu)}{\p s_{\alpha}}\\
\ddot{\nu}_{\iota} &= -\sum_{\beta =1}^{M+C}\frac{\lambda_{\beta}}{\mu_\nu} \frac{\partial \sigma_{\beta}(\vec{R}, s, \nu)}{\partial \nu_\iota}
\end{aligned}
\end{equation}
Let us now consider the limit of zero mass for both sets of auxiliary variables. For this limit to be well defined, it needs to be taken in a homogeneous way. We then set $\mu_{\nu}\equiv \mu_s/{\kappa}$ with $\kappa$ a nonzero constant of dimensions given by the ratio of the masses, so that the last equation in the system above becomes
\begin{equation}
\ddot{\nu}_{\iota} = -\sum_{\beta =1}^{M+C}\kappa \frac{\lambda_{\beta}}{\mu_s} \frac{\partial \sigma_{\beta}(\vec{R}, s, \nu)}{\partial \nu_\iota}
\end{equation}
We can now take the limit $\mu_s\rightarrow 0$ in Eq.~\eqref{eq:MZ_stepone}. In order for the auxiliary variables to have finite acceleration, the ratio $\gamma_\beta = \lim_{\mu_s\to0}\frac{\lambda_\beta}{\mu_s}$ must remain finite, implying that the Lagrange multipliers $\lambda_\beta$ are proportional to $\mu_s$. In the limit of zero mass for the auxiliary variables, then, 
\begin{equation}
\label{eq:BODynamicsPlusNu}
\begin{aligned}
m_i\ddot{\vec{R}}_i &= -\nabla_{\vec{R}_i}V(\vec{R}, s), \\
\ddot{s}_\alpha &= -\sum_{\beta =1}^{M+C}\gamma_{\beta} \frac{\partial \sigma_{\beta}(\vec{R}, s, \nu)}{\partial s_\alpha}, \\
\ddot{\nu}_{\iota} &= -\sum_{\beta =1}^{M+C}\kappa \gamma_{\beta}  \frac{\partial \sigma_{\beta}(\vec{R}, s, \nu)}{\partial \nu_\iota}
\end{aligned}
\end{equation}
Eq.~\eqref{eq:BODynamicsPlusNu} defines the mass-zero constrained dynamics and it enables to recognize most of the interesting properties of the approach mentioned in the Introduction. Firstly, since the Lagrange multipliers $\lambda_\beta$ go to zero with the auxiliary masses $\mu_s$, the evolution of the physical variables does not depend directly on the constraints. Secondly, the dynamics of the $s$ and $\nu$, controlled only by the constraint forces, satisfy by construction all the conditions imposed on the system. This implies that these conditions are automatically fulfilled also in the first of Eqs.~\eqref{eq:BODynamicsPlusNu} which is then equivalent to Eq.~\eqref{eq:BODynamicsPolar}. Thus, by rigorously enforcing the mass-zero limit for the auxiliary variables, the system above provides a classical evolution for all degrees of freedom that leads to the exact adiabatic dynamics for the physical degrees of freedom. Thirdly, the numerical integration of the first equation can be performed with any standard MD algorithms (e.g. Verlet) with a timestep determined only by the force acting on the physical dofs. In addition, at each timestep, the Lagrange multipliers  $\gamma_\alpha$, that appear as unknown, time-dependent parameters in the dynamical system, must be determined. This is done enforcing the constraint, $\sigma_{\alpha}(R(t+dt), s(t+dt), \nu(t+dt))=0$, at the position predicted by the MD algorithm as described in Refs.~\cite{ryckaert:1977,ciccotti:1986}. This approach prevents propagation of the error between values of the variables at different timesteps. In current implementations of the approach, the constraints are satisfied via the SHAKE iterative algorithm, which was proven to be symplectic and time reversible~\cite{leimkuhler:1994,leimkuhler:2004-book}. Note that the homogeneous mass-zero limit has introduced an unknown scaling factor $\kappa$ in the equations. The choice of the numerical value of this parameter depends on the specific system and is discussed more in detail in Refs.~\cite{bonella:2020,coretti:2020a}, where it is also shown that MaZe results are very stable with respect to this choice. 

To conclude this section, note that when no additional constraints are present, i.e. in the absence of the $\nu$ variables, the MaZe system reduces to
\begin{equation}
\label{eq:BODynamics}
\begin{aligned}
m_i\ddot{\vec{R}}_i &= -\nabla_{\vec{R}_i}V(\vec{R}, s), \\
\ddot{s}_\alpha &= -\sum_{\beta =1}^M\gamma_{\beta} \frac{\partial \sigma_{\beta}(\vec{R}, s)}{\partial s_\alpha}
\end{aligned}
\end{equation}
This form of the evolution equations is appropriate, for example, to simulate the shell model and, in view of the specific application considered in the results section and to simplify the notation, we shall adopt it in what follows. In particular, in the next subsection we shall prove that Eq.~\eqref{eq:BODynamics} samples exactly the Born-Oppenheimer probability density for the physical variables, i.e. the last MaZe property mentioned in the Introduction. This is an interesting result because the use of constraints may induce a nontrivial metric in the phase space of the system~\cite{ryckaert:1983,ciccotti:2018} and require appropriate reweighting of statistical properties in the physical phase space. Furthermore, importantly, we shall show that approximate adiabatic separation, i.e. performing a dynamics with $\mu\neq0$, can induce a bias in the statistical properties of the system.

\subsection[Statistical Mechanics of MaZe]{Statistical Mechanics of the Mass-Zero Constrained Evolution}
\label{sec:Statistics}
The discussion in this subsection summarizes the proof presented in Ref.~\cite{coretti:2020a} and is reported here for completeness and for the reader's convenience.
 Let us start by reconsidering the extended system before the mass-zero limit is taken. In the absence of additional conditions, the Lagrangian is given by
 \begin{equation}
\label{eq:Lagrangian}
L(\vec{R}, \dot{\vec{R}}, s, \dot{s}) = \frac{1}{2}\sum_{i=1}^{N}m_i\dot{\vec{R}}^{2}_{i} + \frac{1}{2}\sum_{\alpha=1}^M\mu_s \dot{s}^2_\alpha - V(\vec{R}, s),
\end{equation}
The statistical mechanics of the system is described more naturally using (at first) a convenient set of generalized coordinates and in the Hamiltonian formalism. Proceeding in analogy with Ref.~\cite{ciccotti:2018}, we then start by performing the change of variables
\begin{equation}
\label{eq:ChangeOfVariables}
\begin{aligned}
\vec{R}_i &\mapsto \vec{\rho}_i = \vec{R}_i \\
s_\alpha &\mapsto \sigma_\alpha = \sigma_\alpha(\vec{R}, s)
\end{aligned}
\end{equation}
In the following, we shall use the notation $\upsilon = (\vec{\rho}, \sigma)$ where $\vec{\rho}=\{\vec{\rho}_1,...,\vec{\rho}_{N}\}$ and $\sigma=\{\sigma_1,...,\sigma_M\}$ and observables expressed in the new variables will be denoted in calligraphic font. The Hamiltonian of the system can be obtained via standard Legendre transform of the Lagrangian $\mathcal{L}(\upsilon,\dot{\upsilon})=L\bigl(R(\upsilon),s(\upsilon),\dot{R}(\upsilon,\dot{\upsilon}),\dot{s}(\upsilon,\dot{\upsilon})\bigr)$, and is given by
\begin{equation}\label{eq:Hamiltonian}
\mathcal{H}(\upsilon, \pi^{\upsilon}) =  \frac{1}{2}\pi^{\upsilon \transp} \mathbb{M}^{-1}(\upsilon)\pi^{\upsilon} + \mathcal{V}(\upsilon)
\end{equation}
where the momentum $\pi^{\upsilon}$ is 
\begin{equation}
\begin{aligned}
&\pi^{\upsilon}_k = \frac{\p \mathcal{L}(\upsilon, \dot{\upsilon})}{\p \dot{\upsilon}_k} \\
&=
\begin{cases} 
\vec{\pi}^{\rho}_i = \nabla_{\dot{\vec{\rho}}_i}\mathcal{L}(\vec{\rho}, \dot{\vec{\rho}}, \sigma, \dot{\sigma})& \text{$k = 1,\dots, 3N$}\\
\pi^{\sigma}_\alpha = \frac{\p \mathcal{L}(\vec{\rho}, \dot{\vec{\rho}}, \sigma, \dot{\sigma})}{\p \dot{\sigma}_\alpha}& \text{$k = 3N+1,\dots, 3N+M$ }\\
\end{cases}
\end{aligned}
\end{equation}
($i=1,\dots,N$ and $\alpha = 1,\dots,M$), and we have also introduced $\mathbb{M}^{-1}(\upsilon)$,  i.e. the inverse of the metric matrix 
\begin{equation}
\begin{aligned}
\mathbb{M}_{kk'} = \sum_{i=1}^{N}m_i\frac{\p \vec{R}_i}{\p \upsilon_k}\cdot\frac{\p \vec{R}_i}{\p \upsilon_{k'}} + \sum_{\alpha=1}^{M}\mu_s\frac{\p s_\alpha}{\p \upsilon_k}\frac{\p s_\alpha}{\p \upsilon_{k'}}
\end{aligned}
\end{equation}
associated to the new variables.
For future convenience, the metric matrix and its inverse can also be expressed in block form as
\begin{equation}
\label{eq:MassMatrix}
\mathbb{M} = 
\begin{bmatrix}
\mathbb{A} & \mathbb{B} \\
\mathbb{B}\transp & \mathbb{\Gamma}
\end{bmatrix}
\quad
\text{with inverse}
\quad
{\mathbb{M}}\inv = 
\begin{bmatrix}
\mathbb{\Delta} & \mathbb{E} \\
{\mathbb{E}}\transp & \mathbb{Z}
\end{bmatrix}
\end{equation}
where $\mathbb{A}$ and $\mathbb{\Delta}$ are $3N\times 3N$ matrices, $\mathbb{B}$ and $\mathbb{E}$ are $3N\times M$ matrices, and $\mathbb{\Gamma}$ and $\mathbb{Z}$ are $M\times M$ matrices, whose expressions are given in~\ref{subsec:AppADefMat}. 

The average of an observable $\mathcal{O}(\vec{\rho}, \sigma, \vec{\pi}^{\rho}, \pi^{\sigma})$ in the constrained microcanonical ensemble is given by
\begin{equation}
\label{eq:ConstrainedAverage1}
\begin{aligned}
\Braket{\mathcal{O}} = \frac{1}{\mathcal{Z}}\int&\dd^{3N}\rho\dd^{3N}\pi^{\rho}\dd^M\sigma\dd^M\pi^{\sigma} \delta^M(\sigma)\delta^M(\pi^{\sigma} - \tilde{\pi}^{\sigma})\\
&\times \delta(\mathcal{H}(\vec{\rho}, \sigma, \vec{\pi}^{\rho}, \pi^{\sigma})-E) \mathcal{O}(\vec{\rho}, \sigma, \vec{\pi}^{\rho}, \pi^{\sigma})\\
\end{aligned}
\end{equation}
where $\mathcal{Z}$ is the partition function. The delta functions in the equation above express the constant energy condition (second line) and the constraints (first line). Note that, in addition to the delta function associated to the holonomic constraints, $\delta^M(\sigma)$, the integrand contains a delta function involving the momenta $\pi^{\sigma}$. This delta originates from the fact that, in order for the constraints to be satisfied at all times, the additional condition $\dot{\sigma}=0$ must hold. Using the relation  $\dot{\upsilon}={\mathbb{M}}\inv\pi^{\upsilon}$, this implies (see also~\ref{subsec:AppAIntMom}) that, when the constraints are imposed, the momenta must satisfy
\begin{equation}
\label{eq:ConstrainedMomentum}
\pi^{\sigma}=\tilde{\pi}^{\sigma} = -{\widetilde{\mathbb{Z}}}\inv{\widetilde{\mathbb{E}}}\transp\cdot\vec{\pi}^{\rho}
\end{equation}
where the tildes indicate that all matrices are evaluated at $\sigma=0$. We now move to determine the expression for the average after integration over the variables associated with the constraints. To that end, the integral over $\pi^{\sigma}$ is evaluated first, followed by the change of variables $\sigma_\alpha \mapsto s_\alpha$, $\vec{\rho}_i \mapsto \vec{R}_i$, and a last integration over the $s$ variables. The last two steps are discussed more in detail in Ref.~\cite{bonella:2020} and in~\ref{subsec:AppAIntMom}. Here we report the result of these operations, which is given by
\begin{equation}
\begin{aligned}\label{eq:ZeroMassMargProb}
\Braket{\mathcal{O}} =\frac{1}{\mathcal{Z}'}\int&\dd^{3N} R\dd^{3N}\pi^{\rho} O(\vec{R}, \tilde{s}, \vec{\pi}^{\rho}, \tilde{\pi}^{\sigma})\\ &\times \delta(H(\vec{R}, \tilde{s}, \vec{\pi}^{\rho}, \tilde{\pi}^{\sigma})-E)
\end{aligned}
\end{equation}
In the equation above, $\tilde{s}=\tilde{s}(\vec{R})$ is such that $\sigma(\vec{R},\tilde{s})\equiv0$ (we assume, as commonly done in the Born-Oppenheimer framework that this expression has, for any $\vec{R}$, a single root), and $\tilde{\pi}^{\sigma}$ is defined in Eq.~\eqref{eq:ConstrainedMomentum}. The Hamiltonian is also evaluated on the hypersurface $\sigma=0$, $\pi^{\sigma}=\tilde{\pi}^{\sigma}$ where the constrained motion takes place. Its explicit form is derived in~\ref{subsec:AppAHamCon} and is equal to
\begin{equation}
\label{eq:ConstrainedHamiltonian}
\begin{aligned}
H(\vec{R}, \tilde{s}, \vec{\pi}^{\rho}, \tilde{\pi}^{\sigma} )= \frac{1}{2}
\vec{\pi}^{\rho\transp}\widetilde{\mathbb{A}}\inv\vec{\pi}^{\rho} + V(\vec{R}, \tilde{s}) 
\end{aligned} 
\end{equation}
Eq.~\eqref{eq:ZeroMassMargProb} defines the average of an observable with respect to a marginal probability where the constraints (or equivalently the auxiliary variables) have been integrated over. This marginal probability, however, still depends on the value of the mass, $\mu_s$, associated to the auxiliary dofs. This is apparent in the definition of the mass matrix $\widetilde{\mathbb{A}}$ (and its inverse) and therefore of the generalized momenta $\vec{\pi}^{\rho}$, and from the dependence of the observable on $\tilde{\pi}^{\sigma}$. Indeed, from the definition in~\ref{subsec:AppADefMat}, we have 
\begin{equation}
\widetilde{\mathbb{A}}=\mathbb{D}+\mu_s\widetilde{\mathbb{R}}=\mathbb{D}\left[\mathds{1}+\mu_s\mathbb{D}\inv\widetilde{\mathbb{R}}\right]
\end{equation}
where $\mathbb{D}_{jj'}=m_j\delta_{jj'}$ and $\widetilde{\mathbb{R}}_{jj'}=\sum_{\alpha=1}^M \frac{\p s_{\alpha}}{\p \vec{\rho}_j}\cdot \frac{\p s_{\alpha}}{\p \vec{\rho}_{j'}}$ so that
\begin{equation}
\label{eq:AMatrix}
\widetilde{\mathbb{A}}\inv=\left[\mathds{1}+\mu_s\mathbb{D}\inv\widetilde{\mathbb{R}}\right]\inv\mathbb{D}\inv
\end{equation} 
Furthermore, the relation $\pi^{\upsilon}={\mathbb{M}}\dot{\upsilon}$ implies, $\vec{\pi}^{\rho}=\mathbb{A}\cdot\dot{\vec{\rho}}+\mathbb{B}\dot{\sigma}$ giving, on the constrained hypersurface,
\begin{equation}
\label{eq:MomentaRel}
\vec{\pi}^{\rho}=\widetilde{\mathbb{A}}\cdot\dot{\vec{\rho}}=\widetilde{\mathbb{A}}\cdot\dot{\vec{R}}
\end{equation}
where in the last equality we used $\dot{\vec{\rho}}=\dot{\vec{R}}$, as implied by the change of variable $\vec{R}\mapsto\vec{\rho}$. Finally (see~\ref{subsec:AppADefMat}), on the constrained hypersurface, we also have
\begin{equation}\label{eq:ConstrainedMomentumNew}
\begin{aligned}
\tilde{\pi}^{\sigma} &= -{\widetilde{\mathbb{Z}}}\inv{\widetilde{\mathbb{E}}}\transp\cdot\vec{\pi}^{\rho} \\
&= \left[\widetilde{\mathbb{\Gamma}}-\widetilde{\mathbb{B}}\transp\widetilde{\mathbb{A}}\inv\widetilde{\mathbb{B}}\right]{\widetilde{\mathbb{E}}}\transp\cdot\vec{\pi}^{\rho} 
\end{aligned}
\end{equation}
that also carries a dependence on the mass $\mu_s$ due to the definition of the matrices $\widetilde{\mathbb{\Gamma}}$ and $\widetilde{\mathbb{B}}$. Let us now consider the limit $\mu_s \rightarrow 0$. $\widetilde{\mathbb{\Gamma}}$ and $\widetilde{\mathbb{B}}$ are proportional to $\mu_s$ (see~\ref{app:MaZeStatMech}) and therefore vanish in the limit, while $\lim_{\mu_s\rightarrow 0} \widetilde{\mathbb{A}} = \widetilde{\mathbb{D}}$.
In the zero auxiliary mass limit then, $\tilde{\pi}^{\sigma}=0$ and the Hamiltonian of the system becomes
\begin{equation}
H(\vec{R}, \tilde{s}, \vec{\pi}^{R}, \tilde{\pi}^{\sigma} = 0)=\frac{1}{2}\vec{\pi}^{R\transp}\mathbb{D}\inv\vec{\pi}^{R} + V(\vec{R}, \tilde{s})
\end{equation}
with $\mathbb{D}\inv=\frac{1}{m_j}\delta_{jj'}$ and where we have used the fact that, from Eq.~\eqref{eq:MomentaRel}, in the null auxiliary mass limit, $\vec{\pi}^{\rho}=\mathbb{D}\dot{\vec{R}}=\vec{\pi}^R$.

Substituting in the expression for the average, we obtain
\begin{equation}
\begin{aligned}\label{eq:ZeroMassMargAverage}
\Braket{\mathcal{O}} =\frac{1}{\mathcal{Z}'}\int&\dd^{3N} R\dd^{3N}\pi^{R}O(\vec{R}, \tilde{s}, \vec{\pi}^{R}, \tilde{\pi}^{\sigma}=0) \\ &\times \delta(H(\vec{R}, \tilde{s}, \vec{\pi}^{R}, \tilde{\pi}^{\sigma}=0)-E)
\end{aligned}
\end{equation}

The result above implicitly defines the microcanonical marginal probability in the physical phase space in the full adiabatic limit. This definition is in agreement with the form usually assumed for the Born-Oppenheimer probability. The discussion above also indicates that the dynamical systems rigorously samples this density only in the full $\mu_s\rightarrow 0$ limit and that, for finite auxiliary masses, corrections to the mass matrix associated to the momenta would be required, as indicated by Eqs.~\eqref{eq:AMatrix} and~\eqref{eq:MomentaRel}.

\section{Magnetic MaZe}
\label{sec:MagneticMaZeShell}
In this section we extend the MaZe formalism to treat the shell model~\cite{dick:1958} in constant external magnetic field. The shell model is one of the first attempts to represent polarization effects via empirical potentials, with specific focus to taking into account dipole polarization. 
This is described by assuming that the ion's total charge is divided between a core (representing the nucleus) and a massless shell (representing the electronic charge density). Each core-shell pair is bound by a harmonic potential and feels electrostatic interactions with the other particles. The ions evolve according to full adiabatic dynamics, i.e. subject to a force computed with auxiliary variables at minimum of potential. Although several refinements have been proposed for classical polarizable potentials, the shell model still represents a valid benchmark and was chosen in this work, focused on exploratory calculations for a new development of MaZe, due to its simplicity. Polarization effects are important to capture accurately features of ionic systems ranging from phonon dispersion curves to structural and transport properties~\cite{stoneham:1986}. In the following, we shall consider how they influence charge diffusion in the presence of an external magnetic field. To the best of our knowledge, this is the first time in which this problem is explored with any classical model of polarization and we consider it a first test on the way to more realistic simulations.

Assigning a charge $Q_i$ to core $i$ and $q_{\alpha}$ to the shell $\alpha$, respectively, the MaZe dynamics for the system in magnetic field is conveniently obtained, in analogy with the discussion in Section~\ref{sec:MaZeDynamics}, by first considering the Lagrangian 
\begin{eqnarray}
L(\vec{R}, \dot{\vec{R}}, \vec{S}, &\dot{\vec{S}}&)= \frac{1}{2}\sum_{i=1}^{N}m_i\dot{\vec{R}}^{2}_{i} + \frac{1}{2}\sum_{\alpha=1}^N \mu \dot{\vec{s}}^2_{\alpha}  \\
&-& V(\vec{R}, \vec{S}) +  \sum_{i=1}^{N} Q_i\dot{\vec{R}}_i\cdot \vec{A}(\vec{R}_i) \nonumber \\
&+& \sum_{\alpha=1}^N q_\alpha \dot{\vec{s}}_{\alpha} \cdot \vec{A}(\vec{s}_{\alpha})\nonumber
\end{eqnarray}
where a finite mass $\mu$ has been (temporarily) associated with the shell variables. In the equation above, $V(\vec{R},\vec{S})$ is the total interaction potential, whose form is detailed in~\ref{app:ShellModel}, and we have introduced the notation $\vec{R}=\{\vec{R}_1,...,\vec{R}_N\}$ (with $\vec{R}_i=\{R^x_i,R^y_i,R^z_i\}$), and $\vec{S}=\{\vec{s}_1,...,\vec{s}_N\}$ (with $\vec{s}_{\alpha}=\{s^x_{\alpha},s^y_{\alpha},s^z_{\alpha}\}$) for the $3N$ Cartesian coordinates of the cores and of the $3N$ shell variables, respectively. $\vec{A}(\vec{r})$ is the vector potential associated to the magnetic field $\vec{B} = \vec{\nabla}_{\vec{r}} \times \vec{A}(\vec{r})$ at position $\vec{r}$. We shall consider a system in a constant magnetic field parallel to the $z$ axis: $\vec{B} = (0, 0, B_z)$. In the Coulomb gauge ($\vec{\nabla}_{\vec{r}} \cdot \vec{A}(\vec{r}) = 0$), a valid choice for the vector potential is then $\vec{A}(\vec{r}) = B_z/2(-y, x, 0)$. The dynamics of the system is defined as fully adiabatic: the shells are assumed to adapt instantaneously to the positions of the cores so that the force on each shell variable is null:
\begin{equation}\label{eq:MinAux}
\begin{aligned}
{\sigma}_{\alpha}^x (\vec{R},\vec{S},\dot{\vec{S}})& = - \frac{\partial V(\vec{R},\vec{S})}{\partial s_{\alpha}^x} + q_{\alpha}B_z\dot{s}_{\alpha}^y=0 \\
{\sigma}_{\alpha}^y(\vec{R},\vec{S},\dot{\vec{S}}) & = - \frac{\partial V(\vec{R},\vec{S})}{\partial s_{\alpha}^y} - q_{\alpha}B_z\dot{s}_{\alpha}^x=0 \\
{\sigma}_{\alpha}^z (\vec{R},\vec{S})& = - \frac{\partial V(\vec{R},\vec{S})}{\partial s_{\alpha}^z} =0
\end{aligned}
\end{equation}
In the equations above, $\alpha=1,...,N$ and we have written explicitly the conditions of zero force for the components on the $xy$ plane, which include the Lorentz force, and the component along the $z$ axis, i.e. parallel to the magnetic field. Eq.~\eqref{eq:MinAux} can still be interpreted a set of $3N$ constraints but, for this system, the components of the force on the plane orthogonal to the field depend on the velocity and the corresponding constraints are therefore no longer holonomic (the $\sigma_{\alpha}^z$, on the other hand, are holonomic). The linear dependence on the velocity of these constraints,\footnote{Constraints of this form are often referred to as semiholonomic} however, still enables to write the Lagrangian equations of motion. In fact, for systems with mixed holonomic and semiholonomic constraints, these equations can be expressed as~\cite{goldstein:2002-book,saletan:1970,flannery:2005,flannery:2011a,flannery:2011b,kutteh:1999,kneller:2017}
\begin{equation}\label{eq:LagrangeNonHolo}
\frac{d}{dt}\frac{\partial L}{\partial \dot{\xi}_{\alpha}^h} - \frac{\partial L}{\partial \xi_{\alpha}^h} = - \sum_{\beta =1}^N\sum_{l=x,y} \lambda_{\beta}^l \frac{\partial \sigma_{\beta}^l}{\partial \dot{\xi}_{\alpha}^h} - \sum_{\beta =1}^N \lambda_{\beta}^z \frac{\partial \sigma_{\beta}^z}{\partial \xi_{\alpha}^z}
\end{equation}
where we have indicated all the dynamical variables with the notation $\mathbbm{R}^{3N+3N} \ni \xi = (\vec{R}, \vec{S})$. Following the same steps described in Section~\ref{sec:MaZeDynamics}, the MaZe dynamical system is derived by first obtaining the evolution equations for the system on the basis of Eq.~\eqref{eq:LagrangeNonHolo}, then rearranging the evolution equations for the shell variables exploiting the condition of  null force as in Eq.~\eqref{eq:MZ_stepone}. In the $\mu\rightarrow 0$ limit, the resulting dynamical system then is
\begin{equation}
\label{eq:MaZeNonHolo}
\begin{aligned}
m_i \ddot{R}^h_i &= -\frac{\partial V(\vec{R},\vec{S})}{\partial R^h_i} + Q_i(\dot{\vec{R}}_i \times \vec{B})_h\\ 
\ddot{s}^h_{\alpha} &=  - \sum_{\beta =1}^N\sum_{l=x,y} \gamma_{\beta}^l \frac{\partial \sigma_{\beta}^l}{\partial \dot{s}_{\alpha}^h} - \sum_{\beta =1}^N \gamma_{\beta}^z \frac{\partial \sigma_{\beta}^z}{\partial s_{\alpha}^h} 
\end{aligned}
\end{equation}
where, as in the previous case, $\gamma_{\beta}^h=\lim_{\mu\to0}\lambda_{\beta}^h/\mu$.

\subsection{The MaZe algorithm for the shell model}
\begin{figure*}
\begin{center}
  \includegraphics[width=\textwidth]{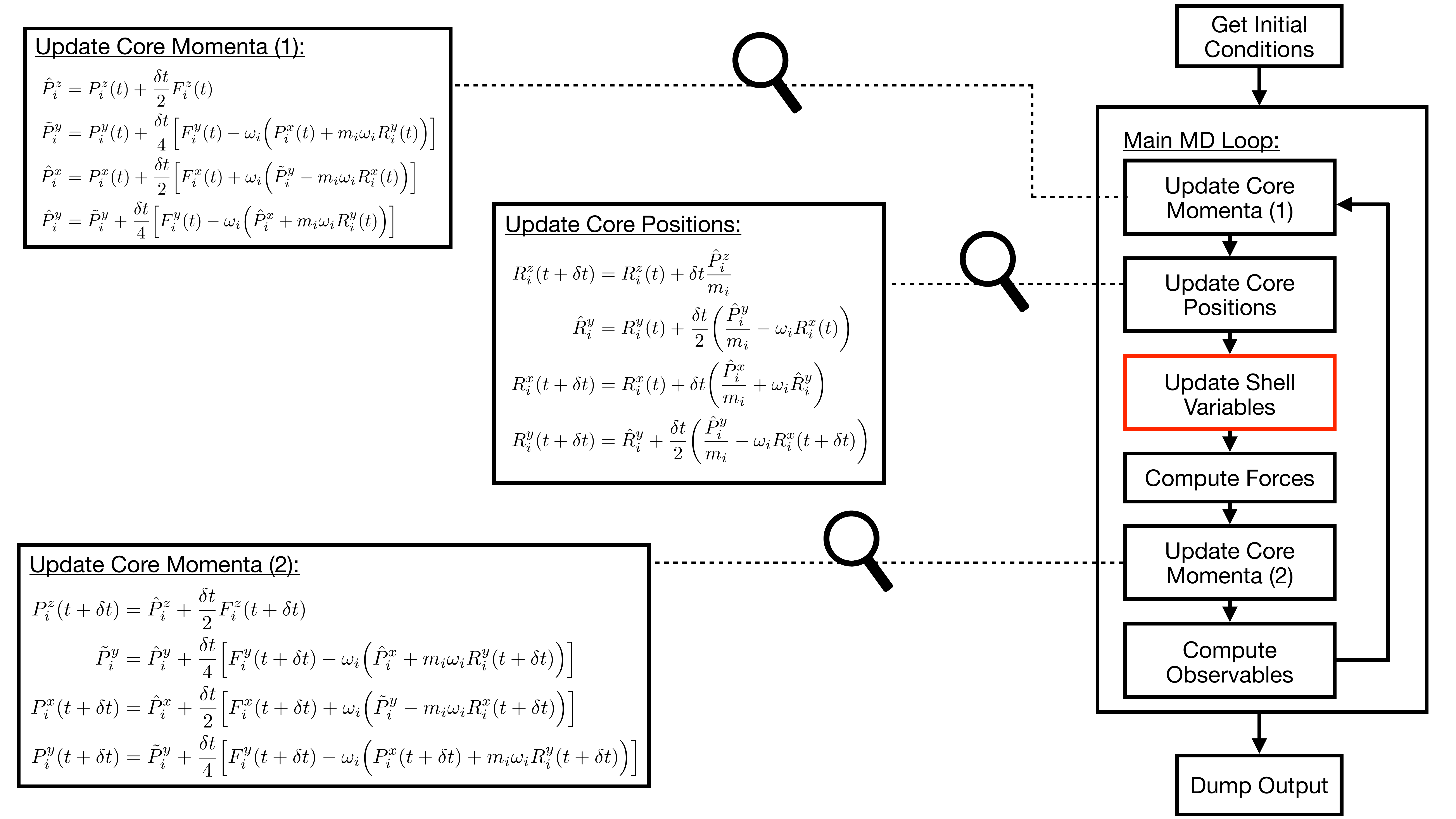}
\end{center}
\caption{On the right: schematic flow chart of the main molecular dynamics loop for the propagation of the cores. The equations for the update of positions and momenta, derived in Ref.~\cite{mouhat:2013}, are given in the boxes on the left. The update of the shell variables, red box in the main loop, is performed via the adapted SHAKE algorithm described in the last part of this section.}
\label{fig:SchemeOfAlgorithm}       
\end{figure*}

The numerical integration of Eqs.~\eqref{eq:MaZeNonHolo} must take into account two nontrivial features: the velocity dependence of the Lorentz force, which prevents direct use of standard integration algorithms (e.g. velocity Verlet), and the presence of nonholonomic constraints. 
These difficulties are solved combining a symplectic algorithm recently introduced to integrate the dynamics of ions in constant external magnetic field~\cite{mouhat:2013}, with an adapted SHAKE algorithm to update the shell's positions and velocities. To set the stage, we introduce the auxiliary dynamical system
\begin{equation}
\label{eq:AuxDynamicalSystem}
\begin{aligned}
\dot{R}_i^x &= \frac{P^x_i}{m_i} + \omega_iR^y_i \\
\dot{R}_i^y &= \frac{P^y_i}{m_i} - \omega_iR^x_i \\
\dot{R}_i^z &= \frac{P^z_i}{m_i} \\
\dot{s}_{\alpha}^x & = p_{\alpha}^x \\
\dot{s}_{\alpha}^y & = p_{\alpha}^y \\
\dot{s}_{\alpha}^z & = p_{\alpha}^z
\end{aligned}
\quad \quad
\begin{aligned}
\dot{P}_i^x &= -\frac{\partial V}{\partial R^x_i} + \omega_i(P^y_i - m_i\omega_iR^x_i) \\
\dot{P}_i^y &= -\frac{\partial V}{\partial R^y_i} - \omega_i(P^x_i + m_i\omega_iR^y_i) \\
\dot{P}_i^z &= -\frac{\partial V}{\partial R^z_i}  \\
\dot{p}_{\alpha}^x  &= G_{\alpha}^x \\
\dot{p}_{\alpha}^y &= G_{\alpha}^y \\
\dot{p}_{\alpha}^z &=G_{\alpha}^z
\end{aligned}
\end{equation}
where $V \equiv V(\vec{R},\vec{S})$ and where we have introduced the notation
\begin{equation}
   G_{\alpha}^h (\vec{R},\vec{S})=  - \sum_{\beta =1}^N\sum_{l=x,y} \gamma_{\beta}^l \frac{\partial \sigma_{\beta}^l}{\partial \dot{s}_{\alpha}^h} - \sum_{\beta =1}^N \gamma_{\beta}^z \frac{\partial \sigma_{\beta}^z}{\partial s_{\alpha}^h}
\end{equation}
Taking the time derivative of all positions, it is immediate to show that the system above is equivalent to Eq.~\eqref{eq:MaZeNonHolo}. A convenient integration algorithm can now be obtained by exploiting the Liouvillian formalism and writing the single timestep evolution operator associated to Eq.~\eqref{eq:AuxDynamicalSystem} as
\begin{equation}
    \mathcal{U}(\dt) = e^{i\dt\mathcal{L}}
\end{equation}
where the Liouvillian at the exponent is defined as
\begin{equation}
    i\mathcal{L}_X = \dot{\vec{X}}\cdot \nabla_{\vec{X}} \nonumber
\end{equation}
with $\vec{X}=\{\vec{R},\vec{P},\vec{S},\vec{p}\}$, and where, for example, $\vec{P}=\{\vec{P}_1,...,\vec{P}_N\}$.
To proceed, $\mathcal{U}(\dt)$ is approximated via the following Trotter splitting 
\begin{equation}\label{eq:Trott}
\mathcal{U}(\dt)\approx e^{i\frac{\dt}{2}\mathcal{L}_{\vec{P}}} e^{i\dt\mathcal{L}_{\vec{R}}} e^{i\dt\mathcal{L}_{\vec{p}}} e^{i\dt\mathcal{L}_{\vec{S}}} e^{i\frac{\dt}{2}\mathcal{L}_{\vec{P}}}
\end{equation}
In the equation above, we have separated, as commonly done~\cite{tuckerman:2010-book}, the differential operators acting on the coordinates $\vec{R}=\{\vec{R}_1,...,\vec{R}_N\}$ and momenta  of the ions and the shells. Note that, in the exploratory calculations presented here, we have employed a mixed Trotter break up, which is symmetric (for the cores' momenta) and simple (for the shell variables and the cores' positions). The overall error in the approximation of the propagator is then of order $\dt^2$. One more observation is necessary to proceed. Focusing on the physical dynamical variables, we have
\begin{equation}
\begin{aligned}
i\mathcal{L}_{\vec{P}} &= \sum^N_{i=1} \dot{P}_i^{x}\frac{\partial}{\partial P_i^x} + \sum^N_{i=1} \dot{P}_i^{y}\frac{\partial}{\partial P_i^y} + \sum^N_{i=1} \dot{P}_i^{z}\frac{\partial}{\partial P_i^z} \\
&\equiv i\mathcal{L}_{\vec{P}^x} + i\mathcal{L}_{\vec{P}^y} + i\mathcal{L}_{\vec{P}^z} \\
i\mathcal{L}_{\vec{R}} &=\sum^N_{i=1} \dot{R}_i^{x}\frac{\partial}{\partial R_i^x} + \sum^N_{i=1} \dot{R}_i^{y}\frac{\partial}{\partial R_i^y} + \sum^N_{i=1} \dot{R}_i^{z}\frac{\partial}{\partial R_i^z} \\
&\equiv i\mathcal{L}_{\vec{R}^x} + i\mathcal{L}_{\vec{R}^y} + i\mathcal{L}_{\vec{R}^z} 
\end{aligned}
\end{equation}
In the absence of an external magnetic field the exponentials of the Liouvillians in Eq.~\eqref{eq:Trott} correspond to simple translation operators for the components of the momenta ($e^{i\frac{\dt}{2}\mathcal{L}_{\vec{P}}}$) and coordinates ($e^{i\dt\mathcal{L}_{\vec{R}}}$). Furthermore, always in the absence of an external magnetic field, the Liouvillians for the different Cartesian components of these variables commute among themselves so, for example, $e^{i\frac{\dt}{2}\mathcal{L}_{\vec{P}}}=e^{i\frac{\dt}{2}\mathcal{L}_{\vec{P}^x}}e^{i\frac{\dt}{2}\mathcal{L}_{\vec{P}^y}}e^{i\frac{\dt}{2}\mathcal{L}_{\vec{P}^z}}$ and the corresponding translations can be applied sequentially. Operating from left to right on the phase-space variables with the full set of translation operators, the velocity Verlet algorithm is recovered. When the (constant) magnetic field is present, however, the Liouvillians corresponding to the $x$ and $y$ components of the dynamical variables no longer commute. For example, by applying the commutator of the Liouvillians to a generic function of $\vec{X}$, it can be seen that
\begin{equation}
    [i\mathcal{L}_{\vec R^x},i\mathcal{L}_{\vec R^y}] = -\sum^N_{i=1}\omega_i\left[ \dot{R}_i^{x}\frac{\partial}{\partial R_i^y} + \dot{R}_i^{y}\frac{\partial}{\partial R_i^x}\right]
\end{equation}
A nonzero result is obtained also for $[i\mathcal{L}_{\vec P^x},i\mathcal{L}_{\vec P^y}]$ while the remaining commutators are zero. Due to this, two more Trotter break ups become necessary to write the single-step propagator as a sequence of translations for the different variables. In particular, following Ref.~\cite{mouhat:2013}, we choose the splitting
\begin{equation}\label{eq:FullTrotter}
\begin{aligned}
e^{i\frac{\dt}{2}\mathcal{L}_{\vec{P}}} e^{i\dt\mathcal{L}_{\vec{R}}} e^{i\frac{\dt}{2}\mathcal{L}_{\vec{P}}} =& e^{i\frac{\dt}{4}\mathcal{L}_{\vec{P}^y}}e^{i\frac{\dt}{2}\mathcal{L}_{\vec{P}^x}}e^{i\frac{\dt}{4}\mathcal{L}_{\vec{P}^y}}e^{i\frac{\dt}{2}\mathcal{L}_{\vec{P}^z}} \\ 
&e^{i\frac{\dt}{2}\mathcal{L}_{\vec{R}^y}}e^{i\dt\mathcal{L}_{\vec{R}^x}}e^{i\frac{\dt}{2}\mathcal{L}_{\vec{R}^y}}e^{i\dt\mathcal{L}_{\vec{R}^z}} \\
& e^{i\dt\mathcal{L}_{\vec{p}}} e^{i\dt\mathcal{L}_{\vec{S}}} \\
& e^{i\frac{\dt}{4}\mathcal{L}_{\vec{P}^y}}e^{i\frac{\dt}{2}\mathcal{L}_{\vec{P}^x}}e^{i\frac{\dt}{4}\mathcal{L}_{\vec{P}^y}}e^{i\frac{\dt}{2}\mathcal{L}_{\vec{P}^z}}
\end{aligned}    
\end{equation}
The actions of these operators can be directly translated into a set of instructions taking the system from time $t$ to time $t+\dt$ and this is detailed in the insets in Figure~\ref{fig:SchemeOfAlgorithm}. In particular, the action of the momentum translations induced by the operators in the first line of the equation above corresponds to the ``Update Core Momenta (1)'' in the figure, while ``Update Core Positions'' shows the effect of the coordinates translations induced by the operators in the second line, and ``Update Core Momenta (2)'' derives from application of the operators in the last line. As indicated in the figure, the evaluation of the forces required to implement the second update of the core's momenta (these forces depend on the shell variables at time $t+\dt$) must be performed under the condition that updated shell velocities and positions satisfy the constraints Eq.~\eqref{eq:MinAux}.
These updated shell variables are determined via a straightforward generalization of the standard SHAKE algorithm, see also Refs.~\cite{kutteh:1999,kneller:2017}, that preserves the key conceptual steps of the method. After the update of the positions of the cores, and based on the propagators associated to $\vec{S}$ and $\vec{p}$ (which is equal to $\dot{\vec{S}}$) in Eq.~\eqref{eq:FullTrotter}, the shell variables are advanced as
\begin{equation}
\label{eq:ProvisionalShellsCoord}
\begin{aligned}
s_{\alpha}^h (t+\dt)&= s_{\alpha}^h(t) + \dt \dot{s}_{\alpha}^h(t) + \frac{\dt^2}{2} \ddot{s}_{\alpha}^h(t) \\
& = s_{\alpha}^h(t) + \dt \dot{s}_{\alpha}^h(t) + \\
&- \frac{\dt^2}{2} \Bigl [\sum_{\beta =1}^N\sum_{l=x,y} \gamma_{\beta}^l \frac{\partial \sigma_{\beta}^l(t)}{\partial \dot{s}_{\alpha}^h} + \sum_{\beta =1}^N \gamma_{\beta}^z \frac{\partial \sigma_{\beta}^z(t)}{\partial s_{\alpha}^h} \Bigr ] \\
& \equiv s_{\alpha}^{h}(t+\dt;\vec{\Gamma})
\end{aligned}
\end{equation}
and 
\begin{equation}
\label{eq:ProvisionalShellsVel}
\begin{aligned}
\dot{s}_{\alpha}^h (t+\dt)& = \dot{s}_{\alpha}^h(t) + \dt \ddot{s}_{\alpha}^h(t) \\
& = \dot{s}_{\alpha}^h(t) + \\
&- \dt \Bigl [\sum_{\beta =1}^N\sum_{l=x,y} \gamma_{\beta}^l \frac{\partial \sigma_{\beta}^l(t)}{\partial \dot{s}_{\alpha}^h}  + \sum_{\beta =1}^N \gamma_{\beta}^z \frac{\partial \sigma_{\beta}^z(t)}{\partial s_{\alpha}^h} \Bigr ] \\
& \equiv \dot{s}_{\alpha}^{h}(t+\dt;\vec{\Gamma})
\end{aligned}
\end{equation}
In the second and third line of Eqs.~\eqref{eq:ProvisionalShellsCoord} and~\eqref{eq:ProvisionalShellsVel}, we have used the form of the shell's acceleration prescribed by the MaZe dynamical system, Eq.~\eqref{eq:MaZeNonHolo}, and it must be noted that the constraint accelerations depend on time via the core and shell positions. Finally, the definitions in the fourth line of Eqs.~\eqref{eq:ProvisionalShellsCoord} and~\eqref{eq:ProvisionalShellsVel} highlight the fact that the updated shell positions and velocities are a function of the Lagrange multipliers represented, for future convenience, as the $3N$ dimensional vector
\begin{equation*}
 \vec{\Gamma}=\{\gamma^x_1, \gamma^y_1,\gamma^z_1,...,\gamma^x_N,\gamma^y_N,\gamma^z_N\}  
\end{equation*}
The value of $\vec{\Gamma}$, at this stage, is yet undetermined. Following the SHAKE strategy, these Lagrange multipliers are computed \emph{a posteriori} imposing that the advanced shell positions and velocities satisfy the constraints at time $t+\dt$:
\begin{equation}
\begin{aligned}
\sigma^l_\alpha\bigl(\vec{R}(t+\dt), \vec{S}(t+\dt;\vec{\Gamma}), \dot{\vec{S}}(t+\dt;\vec{\Gamma})\bigr) & = 0  \\
\sigma^z_\alpha\bigl(\vec{R}(t+\dt), \vec{S}(t+\dt ;\vec{\Gamma})\bigr) & = 0 
\end{aligned}
\end{equation} 
with, as usual, $\alpha=1,...,N$ and $l=x,y$. The expressions above are a system of nonlinear equations for the $\vec{\Gamma}$ that is conveniently solved using the SHAKE algorithm. This is an adapted Newton-Raphson method~\cite{allaire:2007-book} in which the Lagrange multipliers are determined iteratively according to
\begin{equation}\label{eq:GammaUpdate}
\vec{\Gamma}^{(n+1)}=\vec{\Gamma}^{(n)} - \vec{\eta}[\mathbb{J_d}(\vec{S}^{(n)}, \dot{\vec{S}}^{(n)})\inv]
\vec{\Sigma}(\vec{S}^{(n)},\dot{\vec{S}}^{(n)})     
\end{equation}
where the superscript $(n)$ indicates the iteration step. In the equation above, the vector of parameters $\vec{\eta} = (\eta_x, \eta_y, \eta_z)$ was introduced. In standard SHAKE calculations, this vector is not present. However, following a common practice in minimization algorithms~\cite{nocedal:2006-book}, it has been shown~\cite{barth:1995} that using a scaling factor to modulate the magnitude of the SHAKE update can improve convergence. As discussed more in detail in Section~\ref{sec:Results}, for the particular problem considered in this paper, it proved useful to use a different scaling factor for the components of the constraints perpendicular and parallel to the field. This is due to the different nature (holonomic and not) of the constraints shown in Eq.~\eqref{eq:MinAux}. In Eq.~\eqref{eq:GammaUpdate} we have also adopted the notation
\begin{equation*}
 \vec{\Sigma}= \{\sigma^x_1, \sigma^y_1,\sigma^z_1,...,\sigma^x_N,\sigma^y_N,\sigma^z_N\}   
\end{equation*}
to define the vector of the constraints, and $\mathbb{J_d}$ for the $3N\times 3N$ diagonal matrix  
\begin{eqnarray}\label{eq:DiagJac}
  \bigr[\mathbb{J_d}(\vec{S}^{(n)}, \dot{\vec{S}}^{(n)})\bigl]_{ab} &=&\frac{\partial \Sigma_a(\vec{S}^{(n)}, \dot{\vec{S}}^{(n)})}{\partial \Gamma_b}\delta_{ab} \\
  & =& \Bigl [\sum^{3N}_{c=1}\frac{\partial \Sigma_a(\vec{S}^{(n)}, \dot{\vec{S}}^{(n)})}{\partial S_c} \frac{\partial S_c}{\partial \Gamma_b} \nonumber \\ 
  &+& \sum^{3N}_{c=1} \frac{\partial \Sigma_a(\vec{S}^{(n)}, \dot{\vec{S}}^{(n)})}{\partial \dot{S}_c} \frac{\partial \dot{S}_c}{\partial \Gamma_b} \Bigr ]\delta_{ab} \nonumber
\end{eqnarray}
(i.e. the diagonal approximation of the Jacobian matrix of the standard Newton-Raphson method), with $a,b=1,...,3N$. This matrix and the vector of the constraints are updated at each iteration step due to the update in the shell's positions and velocities that is performed according to 
\begin{equation}
\begin{aligned}
\vec{S}&^{(n+1)}=\vec{S}^{(n)} + \dt\dot{\vec{S}}^{(n)} + \\ 
&- \frac{\dt^2}{2} \Bigl [\sum_{\beta =1}^N\sum_{l=x,y} (\gamma_{\beta}^l)^{(n)} \nabla_{\dot{\vec{S}}}\sigma_{\beta}^l(t) + \sum_{\beta =1}^N (\gamma_{\beta}^z)^{(n)} \nabla_{\vec{S}}\sigma_{\beta}^z(t) \Bigr ]  \\
\dot{\vec{S}}&^{(n+1)} =\dot{\vec{S}}^{(n)} + \\
&- \dt \Bigl [\sum_{\beta =1}^N\sum_{l=x,y} (\gamma_{\beta}^l)^{(n)} \nabla_{\dot{\vec{S}}} \sigma_{\beta}^l(t) + \sum_{\beta =1}^N (\gamma_{\beta}^z)^{(n)} \nabla_{\vec{S}}\sigma_{\beta}^z(t) \Bigr ]
\end{aligned}
\end{equation}
The algorithm in Eq.~\eqref{eq:GammaUpdate} can be initialized, for each timestep along the dynamics, with the null vector $\vec{\Gamma}^{(0)} = 0$ or, if it is available, with the value of the Lagrange multipliers computed at the previous timestep. The calculations presented in the following were performed using the sequential update of the components of the vector of constraints first proposed in~\cite{ryckaert:1977}. The iteration process is stopped when the modulus of the largest constraint, i.e. $\rm{max}_a |{\Sigma_a}|$, becomes smaller than a predefined tolerance, typically chosen as close to the numerical precision achievable on the computer.

\section{Simulation set up and results}
\label{sec:Results}
The algorithm detailed in the previous section is applied to compute static and transport properties in a shell model simulation of molten NaCl in external magnetic field. The simulated system contains 108 Na$^+$ and 108 Cl$^-$, placed in a cubic box of side $L= 19.87\um{\angstrom}$, corresponding to a density $\rho = 1.3113\um{g}\um{cm^{-3}}$. Periodic boundary conditions are enforced in all directions. The temperature of the system is set to $T \approx 1350\um{K}$. The specific form of the interatomic potential, $V(\vec{R},\vec{S})$ is given in~\ref{app:ShellModel}, and it is similar to the one adopted in Ref.~\cite{jacucci:1976}. We present results for the polarizable system in the presence and in the absence of a constant magnetic field directed along the $z$ axis. The intensity of the field is chosen so that the magnitude of the Lorentz force on each particle is comparable to that of the forces originating from the other interparticle interactions. As in our previous work on the shell model~\cite{coretti:2018b}, we do not use the method of Ewald sums in the simulations, but rather truncate all interactions at a cut-off radius $r_c = 0.5L$. The truncation of the long-range forces, sometimes adopted in simulations of large systems or when accuracy on the energy is not critical~\cite{wolf:1999,zahn:2002,fennell:2006}, was enforced for convenience. Our calculations are intended as a proof-of-principle validation of the MaZe dynamics in magnetic field and, while qualitative trends in the observables will be described, we are not focused on a realistic description of the system. Note that incorporating Ewald sums in the algorithm does not pose a conceptual problem nor has a significant effect on the numerical cost of the approach, as shown in Ref.~\cite{coretti:2020a} where a state-of-the-art classical model of polarization was considered. 

\begin{figure*}
  \includegraphics[width=\textwidth]{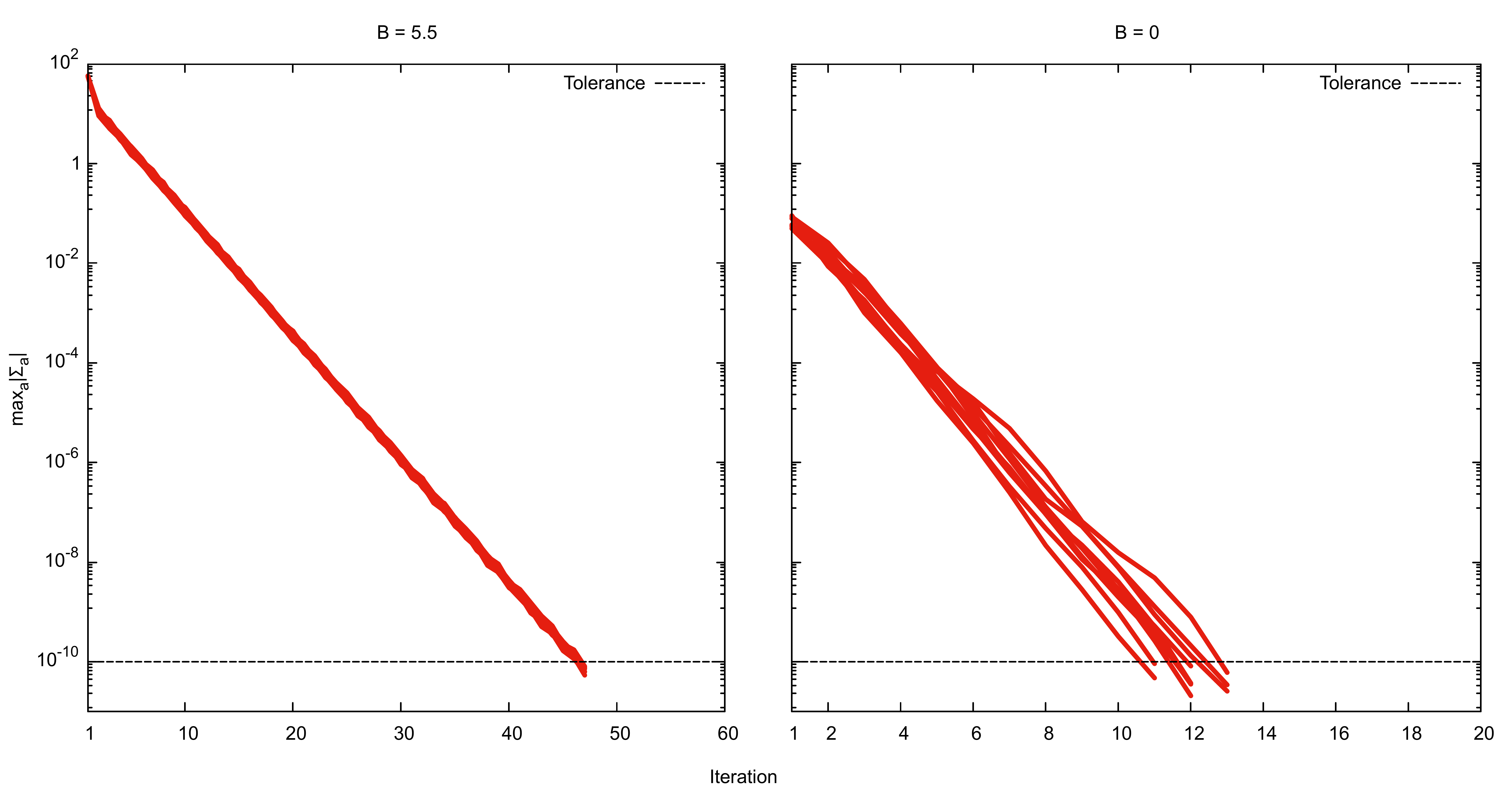}
\caption{Path to convergence of the maximum magnitude of the constraints for typical core configurations along the trajectory. In the right panel, results are shown for the polarizable model in the absence of external magnetic field. In the left panel, results for simulations with $B_z\neq 0$ are reported. The value of the field is indicated in code units in this and in the following figures and it corresponds to $B_z = 8.11 \cdot 10^6$\um{T} in SI units. This large value of $\vec{B}$ is not uncommon in molecular dynamics simulations of particle systems in external magnetic field (see, for example, Refs.~\cite{mouhat:2013,gagliardi:2016,coretti:2021}).}
\label{fig:PathToConvergence}       
\end{figure*}

The simulations are initialized as follows. The Cl$^-$ and Na$^+$ cores are placed on the sites of a simple cubic lattice and then displaced by a small uniform random amount in all directions. Initial shell positions are then found via a conjugate-gradient minimization of the potential energy with respect to these degrees of freedom. Velocities for all the degrees of freedom are set to zero. After calculation of the interatomic forces, the first half of the evolution algorithm is applied to the cores in order to compute $\vec{R}(\delta t)$ and $\vec{P}(\delta t/2)$. New values of the shell variables are determined by applying SHAKE starting from the shell positions found by the conjugate-gradient minimization at step zero. Finally, interatomic forces are computed for this new configuration and the second half of the evolution algorithm is applied to the cores to obtain $\vec{P}(\delta t)$. Note that, because the shell velocities are set to zero at initialization, this procedure ensures that constraints are satisfied at $t=0$ both in the absence and in the presence of the magnetic field. MaZe integration as described in Section~\ref{sec:MaZeDynamics} (no magnetic field) and Section~\ref{sec:MagneticMaZeShell} (magnetic field present) is then started. The timestep for the standard MaZe simulations is set to $\dt=1\um{fs}$, while for the runs with $B_z\neq 0$, $\dt=0.25\um{fs}$ (see below for a discussion of the reasons for the smaller timestep). Equilibration to the target temperature is achieved in all runs by simulating the system for $5 \cdot 10^3\um{fs}$. During this equilibration, the velocities are rescaled if the temperature differs from the target more than $10\%$. NVE runs of total length of $10\um{ps}$ are then performed to compute the properties reported in the following. A strict convergence criterion for the constraints is enforced in all runs by imposing that the maximum magnitude of the constraints is less than $10^{-10}$ units of force. The relaxation parameter $\vec{\eta}$, see Eq.~\eqref{eq:GammaUpdate} and discussion in the previous section, is set to $\vec{\eta}=(1,1,1)$ for the calculations with $B_z=0$ and to $\vec{\eta}=(0.33,0.33,1)$ when the magnetic field is present. While the value of $\vec{\eta}$ can be set via an automatic search and adapted during the run~\cite{barth:1995}, here it was chosen via manual search by optimizing the number of iterations necessary to converge SHAKE for a typical configuration of the cores. Previous experience, confirmed by the simulations reported here, has shown that this is sufficient to provide a stable number of iterations along the whole trajectory. In Figure~\ref{fig:PathToConvergence}, we show the convergence paths of the constraints in the presence (left panel) and absence (right panel) of the magnetic field for a few randomly chosen configurations along the trajectory of the cores. The figures show the magnitude of the largest constraint as a function of the number of SHAKE iterations. In agreement with previous calculations, the number of iterations needed to converge in MaZe calculations for fully holonomic constraints (no magnetic field) is very small and similar for different configurations. The path shows monotonic convergence, with a single slope on the semilogarithmic scale employed in the figure. This fast convergence is facilitated by the already small value of the largest constraint at the start of the iterative process, indicating that the provisional values for the shell positions are quite close to the minimizers of the potential. The convergence of MaZe for the mixed set of holonomic and semiholonmic constraints, on the other hand, is about four times slower. This may be related to the fact that the magnetic force results in larger nondiagonal terms in the Jacobian matrix of the Newton-Raphson procedure, implying that Eq.~\eqref{eq:DiagJac} provides a less effective approximant of its inverse. Furthermore, the value of the maximum constraint at the beginning of the minimization is now larger than in the holonomic case, indicating that our provisional shell positions and velocities at the zero-th iteration are farther from the final solution. This behaviour is sensitive to the choice of the timestep $\dt$, with worse performance (and eventually lack of convergence) with larger timesteps. This suggests that the basin of convergence of SHAKE in the presence of semiholonomic constraints may be smaller than the one for standard applications, an issue that will be further investigated in future studies. Finally, the paths to convergence now present a double slope pattern: a fast initial decay is followed by a slower decrease. Closer inspection of the decrease of individual constraints suggests that this is due to the different speed of convergence of the holonomic and nonholonomic constraints, with the latter evolving faster towards the threshold. This is most likely also related to the different values for the components in the scaling vector $\vec{\eta}$. In spite of the differences in the convergence pattern, and of the need to further investigate the behaviour of the new algorithm, this first implementation of SHAKE for mixed constraints performs well for the nontrivial interactions of the model. 

\begin{figure*}
  \includegraphics[width=\textwidth]{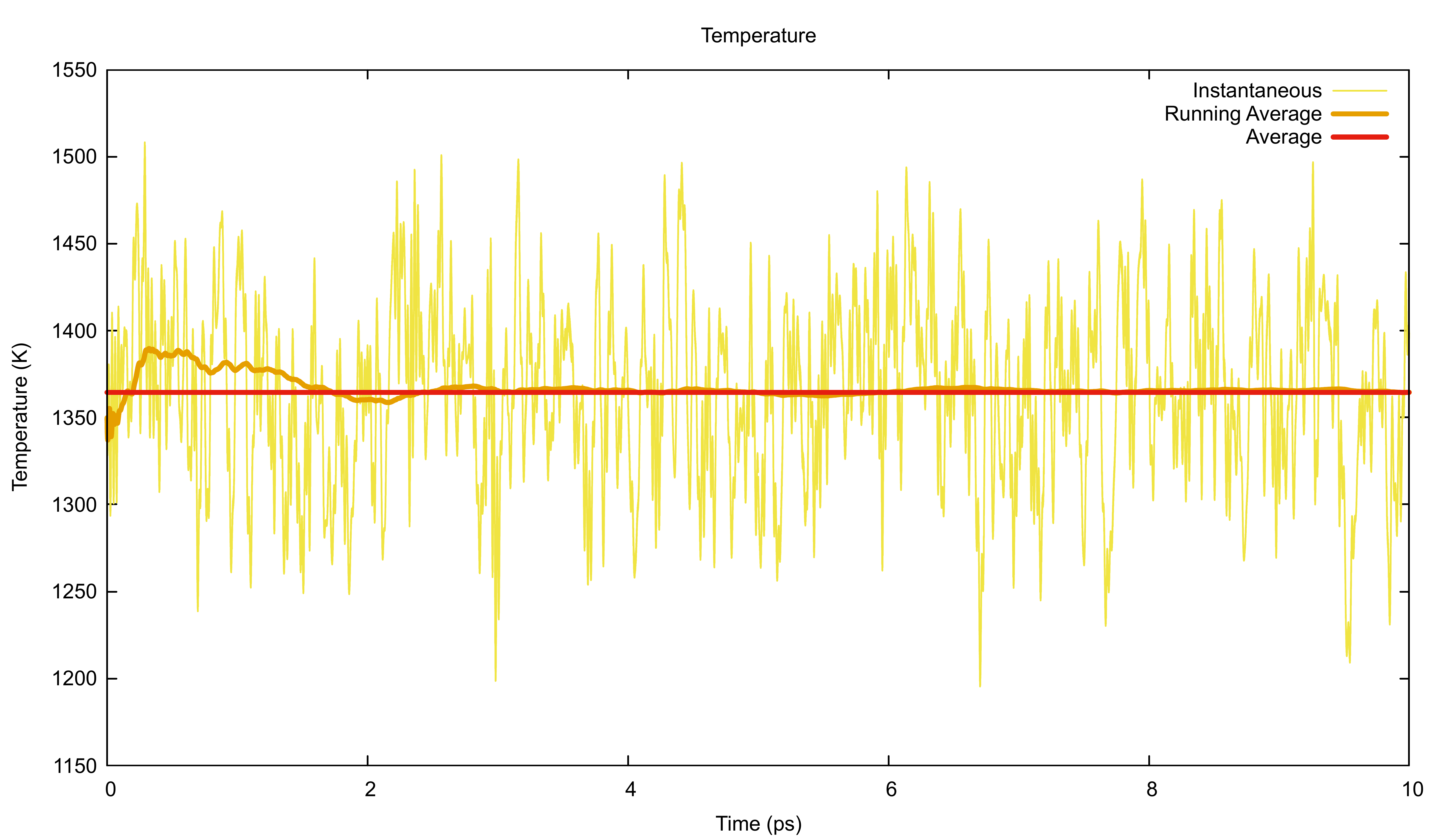}
\caption{Temperature fluctuations (yellow curve), running average (orange line) and final average (red line) along the core trajectory for a system of polarizable ions in magnetic field.}
\label{fig:Temperature}       
\end{figure*}

\begin{figure*}
  \includegraphics[width=\textwidth]{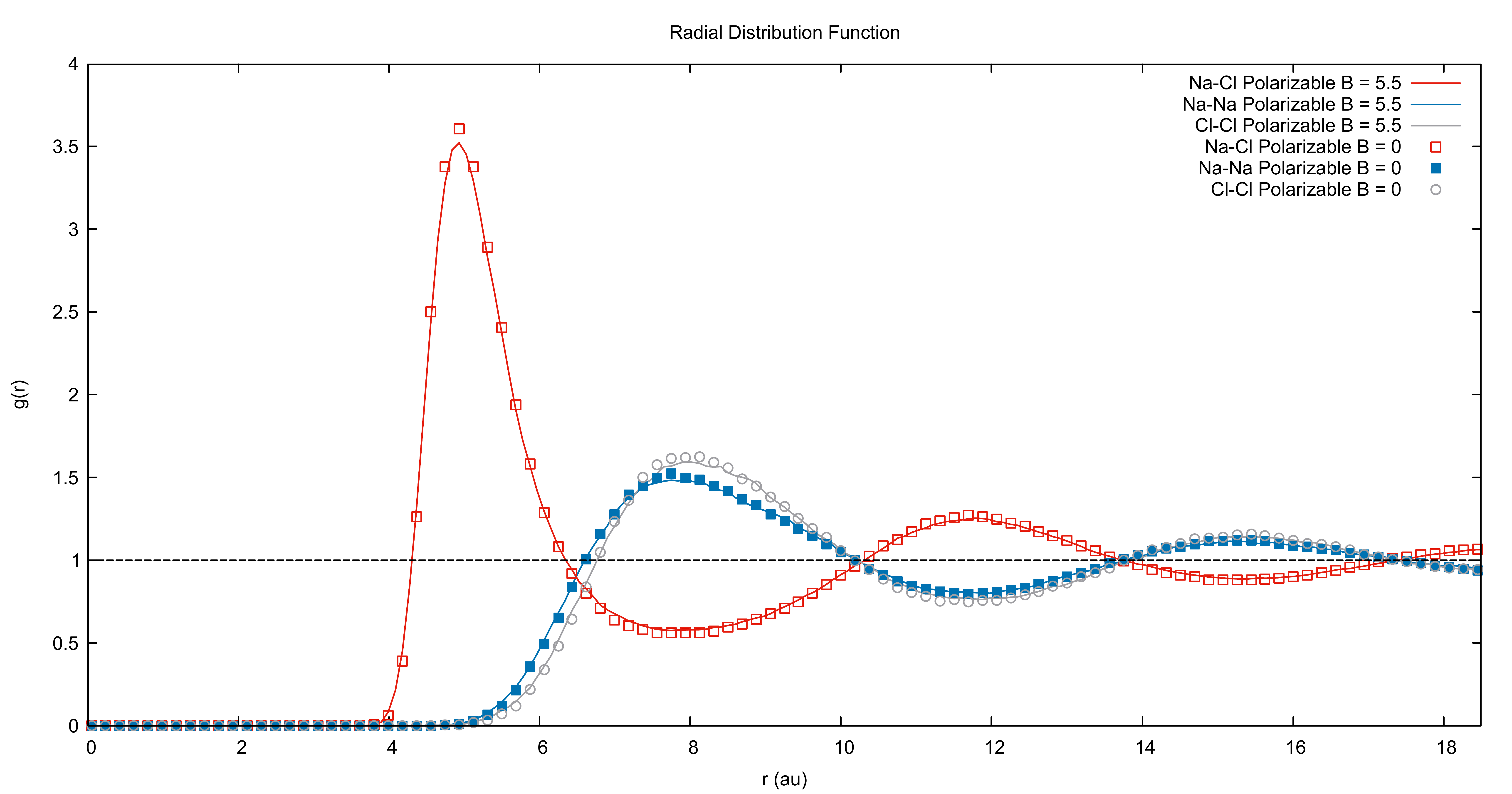}
\caption{Radial distribution function for the ionic species. Solid lines indicate results for the simulations with no magnetic field, while open circles refer to simulations with $B_z\neq 0$.}
\label{fig:RadialDistributionFunctions}       
\end{figure*}

Inspection of typical dynamical indicators and structural properties confirms the reliability of the MaZe approach for classical polarizable models in external magnetic field. In particular, the fluctuations of the total energy relative to the fluctuations of the potential energy along the trajectory of the cores are $\Delta E/\Delta V \approx 2 \cdot 10^{-4}$, where, for example, $\Delta E = \sqrt{\Braket{E^2}-\Braket{E}^2}$. The stability of the new algorithm is visible also in the calculation of the instantaneous temperature of the cores. Figure~\ref{fig:Temperature} shows the fluctuations of this quantity in a $10\um{ps}$ simulation in the presence of the magnetic field, which are --- again --- perfectly compatible with typical results for classical simulations. As a test of the reliability of MaZe semiholonomic dynamics, we consider the radial distribution function of the ionic species as obtained in the simulation with and without external magnetic field. Results are shown in Figure~\ref{fig:RadialDistributionFunctions}, where we report as solid lines the output of the runs in the absence of the magnetic field and as open circles that of the calculations with the magnetic field. The position and shape of the peaks for all $g(R)$ are in good agreement with experimental results~\cite{galamba:2007} and with previous calculations~\cite{mouhat:2013,coretti:2018b} in spite of the somewhat crude treatment of the electrostatic interactions, of a different temperature ($T\approx 1350$ K in this work and $T\approx 1550$ K in Refs.~\cite{mouhat:2013,coretti:2018b}) and of some differences in the parameters in the shell model detailed in~\ref{app:ShellModel}. Perhaps more importantly for our purposes, the curves and the symbols are superimposed. This provides strong validation for the MaZe algorithm presented in Section~\ref{sec:MagneticMaZeShell}: it is in fact known (see, for example, Ref.~\cite{mouhat:2013}) that time-independent averages are not affected by the presence of the magnetic field. As shown by the results on a nontrivial observable, this property is respected by the MaZe algorithm.

\begin{figure*}
  \includegraphics[width=\textwidth]{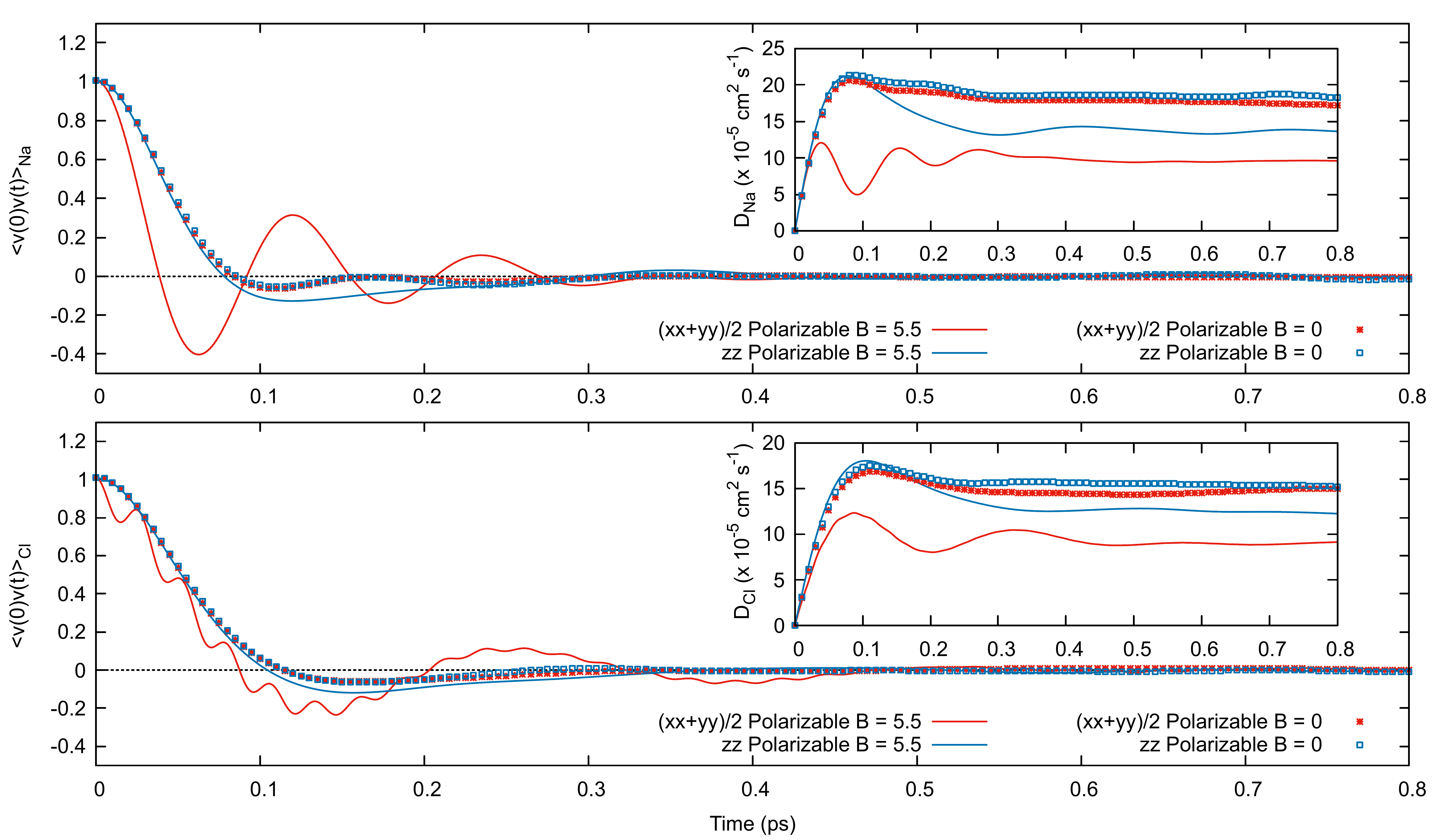}
\caption{Autocorrelation functions of the Cartesian components of the velocity for Na$^+$ (upper panel) and Cl$^-$ (lower panel) for polarizable shell model. Solid lines report results for $B_z \neq 0$, while the open symbols refer to the same system when no magnetic field is present. Diffusion coefficients, computed as time integrals of the correlation functions, are shown in the insets as a function of the total integration time.}
\label{fig:NaClVelCorrDiag}       
\end{figure*}

\begin{figure*}
  \includegraphics[width=\textwidth]{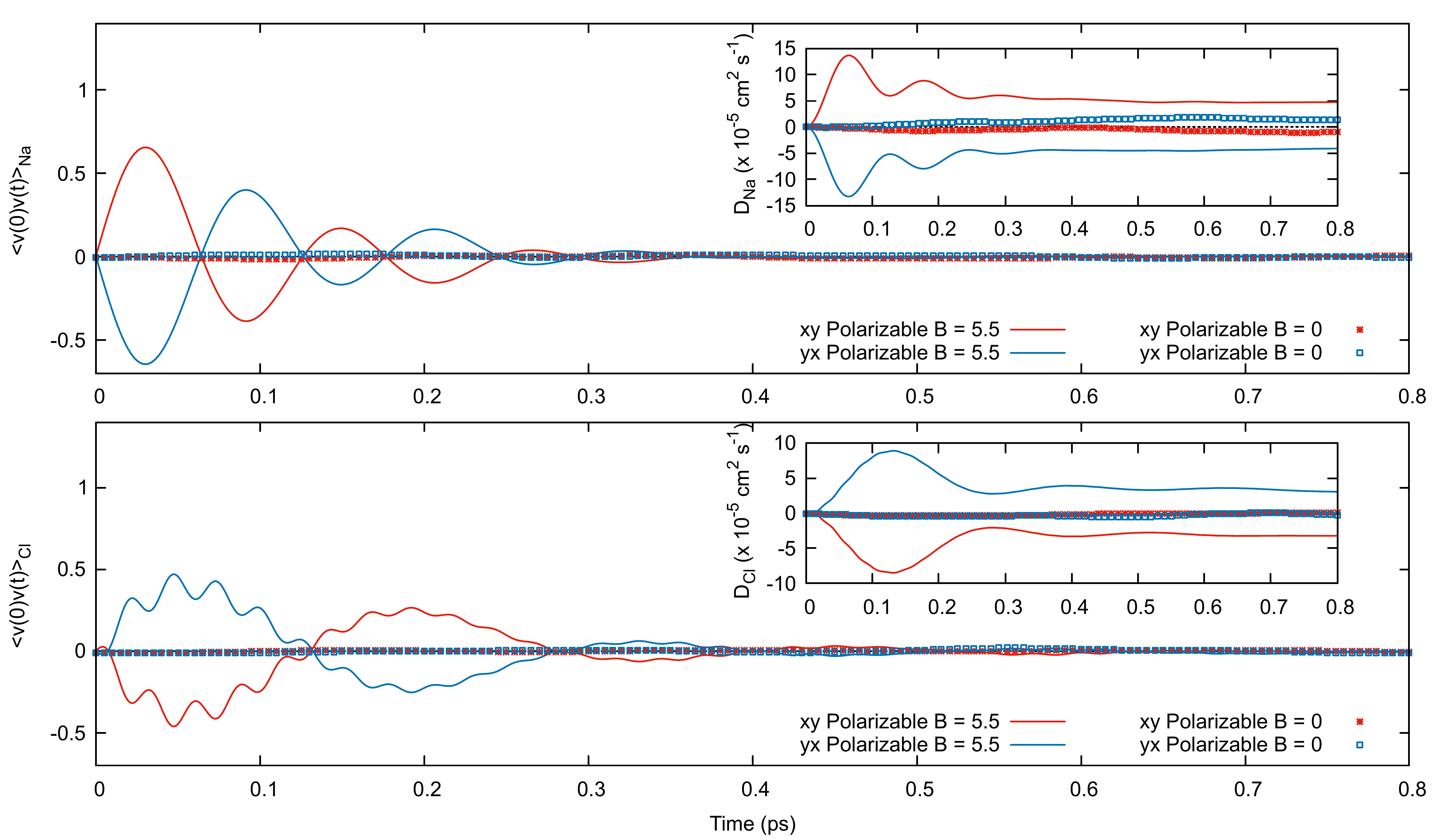}
\caption{$xy$ components of the velocity correlation tensor for Na$^+$ (upper panel) and Cl$^-$ (lower panel) for polarizable shell model. Solid lines report results for $B_z \neq 0$, while the open symbols refer to the same system when no magnetic field is present. Diffusion coefficients, computed as time integrals of the correlation functions, are shown in the insets as a function of the total integration time.}
\label{fig:NaClVelCorrOffDiag}       
\end{figure*}

We now move to the calculation of time-dependent statistical properties of the system. In this case, the presence of a magnetic field is expected to affect the results in nontrivial ways providing further and more interesting testing ground for our approach. We consider, in particular, the velocity correlation functions of the ionic species. In Figure~\ref{fig:NaClVelCorrDiag} and Figure~\ref{fig:NaClVelCorrOffDiag}, we show results for the diagonal and off-diagonal correlations in the absence and presence of $\vec{B}$. Results for system in the absence of the magnetic field are reported in open symbols (Na$^+$ upper panel and Cl$^-$ lower panel). In the insets we also show the corresponding elements of the diffusion tensor. As expected, when $B_z=0$, the three diagonal components of the velocity correlation function are equal for each species and show the characteristic initial decay followed by one minimum or more before going to zero at longer times. For this system, the integral of the autocorrelation function of the velocity yields diffusion coefficients equal to $D_{Na} = (1.78 \pm 0.08) \cdot 10^{-4}\um{cm^2}\um{s^{-1}}$ and $D_{Cl} = (1.51 \pm 0.06) \cdot 10^{-4}\um{cm^2}\um{s^{-1}}$, both obtained averaging the three components on the diagonal of the diffusion tensor. The error on the values of the diffusion is estimated from the off-diagonal components of the tensor for $B_z = 0$. These off-diagonal components must be zero based on time-reversal symmetry arguments~\cite{bonella:2017a,coretti:2018a}, and the results of the simulation can be used to estimate the statistical noise in computing the integrals. All the results discussed above are compatible with previous studies~\cite{galamba:2007} performed on the same system at zero magnetic field.

The presence of the magnetic field breaks the isotropy of space and this implies that the components of the correlation on the plane orthogonal to the field are now different from that in the direction parallel to it, and show an oscillatory behaviour that reflects the rotatory motion induced by the Lorentz force. Consistently, the diffusion coefficients are also affected by the presence of the magnetic field. The observed reduction of their values is in fact a known phenomenon, the so-called magnetoresistance, which is commonly observed for the electrons in semiconductors in the presence of magnetic field and was also reported for ions in previous simulations~\cite{mouhat:2013,gagliardi:2016}. In particular, the diffusion coefficients are reduced to $D^{\perp}_{Na} = (0.96 \pm 0.10) \cdot 10^{-4}\um{cm^2}\um{s^{-1}}$, $D^{zz}_{Na} = (1.36 \pm 0.14) \cdot 10^{-4}\um{cm^2}\um{s^{-1}}$ for the Sodium ions and to $D^{\perp}_{Cl} = (0.91 \pm 0.07) \cdot 10^{-4}\um{cm^2}\um{s^{-1}}$, $D^{zz}_{Cl} = (1.24 \pm 0.10) \cdot 10^{-4}\um{cm^2}\um{s^{-1}}$ for the Chlorine ions, where $D^{\perp} = (D^{xx} + D^{yy})/2$. The effect of the magnetic field is even more striking when considering the off-diagonal components of the velocity correlation functions. In Figure~\ref{fig:NaClVelCorrOffDiag}, we present results for the $xy$ and $yx$ cross-correlations. In the absence of the field, standard time-reversal invariance leads to null values of these quantities. On the other hand, when $B_z\neq 0$ a characteristic oscillatory pattern is observed. As detailed in Ref.~\cite{coretti:2018a}, the antisymmetry of these two observables is dictated by their properties under generalized time-reversal symmetries and well reproduced by our simulations (all other off-diagonal components remain zero, for symmetry reasons). The behaviour of the correlation function is reflected in the values obtained for the $xy$ and $yx$ components of the diffusion tensor, which are now equal to $D^{xy}_{Na} = (0.46 \pm 0.14) \cdot 10^{-4}\um{cm^2}\um{s^{-1}}$ and $D^{yx}_{Na} = (-0.43 \pm 0.14) \cdot 10^{-4}\um{cm^2}\um{s^{-1}}$ for Sodium and to $D^{xy}_{Cl} = (-0.32 \pm 0.10) \cdot 10^{-4}\um{cm^2}\um{s^{-1}}$ and $D^{yx}_{Cl} = (0.32 \pm 0.10) \cdot 10^{-4}\um{cm^2}\um{s^{-1}}$ for Chloride. 

\begin{figure*}
  \includegraphics[width=\textwidth]{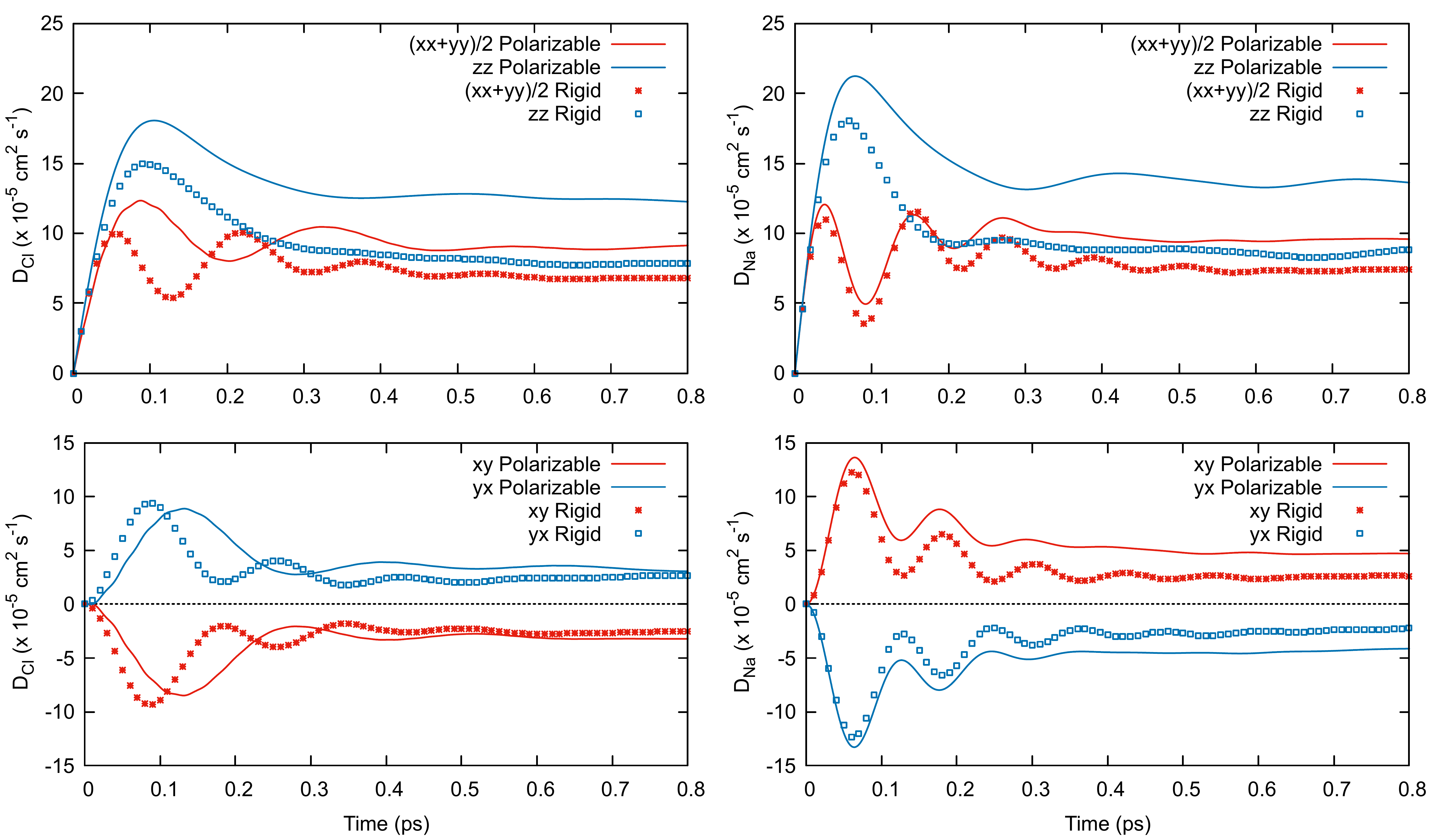}
\caption{Diagonal (upper panels) and off-diagonal, i.e. perpendicular to the magnetic field (lower panels), components of the diffusion tensor for a rigid ion (symbols) and a shell model (lines) system in magnetic field.}
\label{fig:PolNoPolDiff}       
\end{figure*}

Finally, it is interesting to explore the effects of polarization on the transport properties of this model of molten NaCl. To assess the relevance of these effects, we compare the elements of the diffusion tensor discussed above with those from a simulation of an unpolarized model of the system. The unpolarized, or rigid ion, model is defined by removing the shell variables from the electrostatic interactions, while keeping the remaining parameters of the potential for the cores as in our previous runs. Further details and the specific form of the interaction are also given in~\ref{app:ShellModel}. In Figure~\ref{fig:PolNoPolDiff}, we show results for the diffusion coefficients of the polarized (solid curves) and rigid ions (open symbols) simulations in the presence of the same magnetic field. The plot indicates that, for the rigid ion model, the diagonal components of the diffusion tensor (upper panel of the figure) are very similar for both species. The presence of polarization globally enhances transport in the system, but --- due to the different polarizability of the two ions --- leads to different values, in particular, of the component of the diffusion tensor parallel to the magnetic field. Similarly, the cross components of the diffusion for Na$^+$ and Cl$^-$ in the plane orthogonal to the magnetic field (bottom panel of the figure) are essentially identical for rigid ions, but polarization separates them. In particular, for the rigid ion case, we obtain values for the $xy$ and $yx$ components of the diffusion coefficients given by $\bar{D}^{xy}_{Na} = (0.25 \pm 0.14) \cdot 10^{-4}\um{cm^2}\um{s^{-1}}$ and $\bar{D}^{yx}_{Na} = (-0.24 \pm 0.14) \cdot 10^{-4}\um{cm^2}\um{s^{-1}}$ for Sodium and $\bar{D}^{xy}_{Cl} = (-0.26 \pm 0.10) \cdot 10^{-4}\um{cm^2}\um{s^{-1}}$ and $\bar{D}^{yx}_{Cl} = (0.25 \pm 0.10) \cdot 10^{-4}\um{cm^2}\um{s^{-1}}$ for Chloride, where the symbol $\bar{D}$ is used to indicate that the diffusion coefficient is computed for the rigid ion model, at difference with the notation $D$ which indicates diffusion coefficient computed for the shell model.

This has an interesting implication for the detection of the ionic Hall effect in molted NaCl. In the Nerst-Einstein approximation, in fact, the key indicator of this phenomenon, i.e. the Hall mobility, is given by~\cite{gagliardi:2016}
\begin{equation}
\mu_H = \frac{1}{B_z}\frac{D_{Na}^{xy}+D_{Cl}^{xy}}{D_{Na}^{\perp}+D_{Cl}^{\perp}}
\end{equation}
When the off-diagonal components of the diffusion tensor of the two species are equal and opposite, as  in previous more refined calculations on a rigid ion model for the system~\cite{mouhat:2013,gagliardi:2016} and, within errors, in the results shown in Figure~\ref{fig:PolNoPolDiff}, the mobility is obviously null. In particular, the values of the diffusion coefficients obtained from the rigid ion simulations performed in this work yield a value for the Hall mobility given by $\bar{\mu}_H \approx -0.9\cdot10^{-5}\um{cm^2}\um{V^{-1}}\um{s^{-1}}$. On the other hand, the diffusions obtained with the shell model result in $\mu_H \approx 8.5\cdot10^{-5}\um{cm^2}\um{V^{-1}}\um{s^{-1}}$. While error bars (not reported) are still quite large with our level of statistics, the noticeably different values for these mobilities suggest that Hall effect is absent or hardly measurable for the rigid ion model but quite appreciable when polarization is accounted for. This observation needs to be confirmed via more accurate calculations, but it clearly underlines the relevance of polarization on observables affected by relatively subtle effects in transport processes.

\section{Conclusions}
In this paper, we described the MaZe dynamics for the simulation of systems where the evolution of a set of physical dofs depends on parameters subject to assigned conditions. Fully adiabatic dynamics, in \textit{first principle} or classical polarizable models, is perhaps the most relevant example of such systems. The derivation of MaZe and of its key properties was presented, using the classical shell model for polarization as a reference case. MaZe exploits the Lagrangian formulation of classical mechanics to define an extended system in which the external parameters evolve as auxiliary dynamical variables of zero mass. These variables are subject to constraints that strictly enforce the conditions on the parameters for each configuration of the physical dofs in the evolution. The mass-zero value for the auxiliary variables results in rigorous fully adiabatic evolution for the physical dofs and, consequently, on exact statistical sampling of the associated probability density. From a numerical point of view, the integration of the constrained dynamics is efficiently performed using the SHAKE algorithm in its standard form for holonomic systems.

A new development extending this approach to the physically interesting case of the shell model in external magnetic field was also presented. In this case, the presence of the Lorentz force requires to generalize the MaZe formalism and the associated algorithm to systems with constraints that depend linearly on the velocities. This generalization was described in the second part of the paper and used in illustrative calculations on a shell model of molten NaCl. These calculations demonstrate the effectiveness of the new algorithm and provide interesting qualitative information on the effect of polarization on ionic transport in magnetic field. In particular, we showed indications that --- within the model adopted --- polarization is critical to obtain a nonnull value of the Hall mobility for the system. 

\begin{acknowledgements}
The authors are grateful to Rodolphe Vuilleumier, Benjamin Rotenberg e Mathieu Salanne for enlightening discussions. Jean-Paul Ryckaert also deserves special thanks and credit for his role in the birth of the zero mass constrained scheme.
\end{acknowledgements}

\appendix
\section{Addendum to Section~\ref{sec:Statistics}}
\label{app:MaZeStatMech}
\subsection{\textit{Definition of the submatrices in the metric matrix and its  inverse}}\label{subsec:AppADefMat}
\begin{equation}
\label{eq:matrix_direct}
\begin{aligned}
\mathbb{A}_{jj'} &= \sum_{i=1}^{N} m_i \frac{\p \vec{R}_i}{\p \vec{\rho}_j}\cdot\frac{\p \vec{R}_i}{\p \vec{\rho}_{j'}} + \sum_{\alpha=1}^M \mu_s \frac{\p s_{\alpha}}{\p \vec{\rho}_j} \frac{\p s_{\alpha}}{\p \vec{\rho}_{j'}} \\
&= m_j\delta_{ j j'}+ \sum_{\alpha=1}^M \mu_s \frac{\p s_{\alpha}}{\p \vec{\rho}_j} \frac{\p s_{\alpha}}{\p \vec{\rho}_{j'}} \quad \text{for $ j, j' = 1,\dots, N$} \\
\mathbb{B}_{j\beta} &= \sum_{i=1}^{N} m_i \frac{\p \vec{R}_i}{\p \vec{\rho}_j}\cdot\frac{\p \vec{R}_i}{\p \sigma_{\beta}} + \sum_{\alpha=1}^M \mu_s \frac{\p s_{\alpha}}{\p \vec{\rho}_j} \frac{\p s_{\alpha}}{\p \sigma_{\beta}} \\
&= \sum_{\alpha=1}^M \mu_s \frac{\p s_{\alpha}}{\p \vec{\rho}_j} \frac{\p s_{\alpha}}{\p \sigma_{\beta}} \quad \text{for $j = 1,\dots, N$, $\beta =  1,\dots,M$} \\
\mathbb{\Gamma}_{\beta\beta'} &= \sum_{i=1}^{N}m_i\frac{\p \vec{R}_i}{\p \sigma_\beta}\cdot\frac{\p \vec{R}_i}{\p \sigma_{\beta'}} + \sum_{\alpha=1}^{M}\mu_s \frac{\p s_\alpha}{\p \sigma_\beta}\frac{\p s_\alpha}{\p \sigma_{\beta'}} \\
&= \sum_{\alpha=1}^{M}\mu_s\frac{\p s_\alpha}{\p \sigma_\beta}\frac{\p s_\alpha}{\p \sigma_{\beta'}} \quad \text{for $\beta, \beta' =  1,\dots,M$} \\
\end{aligned}
\end{equation}
\begin{equation}
\label{eq:matrix_inverse}
\begin{aligned}
\mathbb{\Delta}_{jj'} &=\sum_{i=1}^{N} \frac{1}{m_i} \nabla_{\vec{R}_i} \vec{\rho}_j\cdot\nabla_{\vec{R}_i} \vec{\rho}_{j'} \quad \text{for $ j, j' = 1,\dots, N$} \\
\mathbb{E}_{j\beta} &= -\sum_{i=1}^{N} \frac{1}{m_i} \nabla_{\vec{R}_i} \vec{\rho}_j \cdot \nabla_{\vec{R}_i}\sigma_{\beta} \quad \text{for $j = 1,\dots, N$,} \\&
\,\,\,\,\,\,\,\,\,\,\,\,\,\,\,\,\,\,\,\,\,\,\,\,\,\,\,\,\,\,\,\,\,\,\,\,\,\,\,\,\,\,\,\,\,\,\,\,\,\,\,\,\,\,\,\,\,\,\,\,\,\,\,\,\,\,\,\,\,\,\,\,\,\,\,\,\,\, \quad \text{$\beta =  1,\dots,M$} \\
\mathbb{Z}_{\beta\beta'} &= \sum_{i=1}^{N} \frac{1}{m_i}\nabla_{\vec{R}_i}\sigma_\beta\cdot\nabla_{\vec{R}_i}\sigma_{\beta'} + \sum_{\alpha=1}^{M}\frac{1}{\mu_s} \frac{\p \sigma_\beta}{\p s_\alpha}\frac{\p \sigma_{\beta'}}{\p s_\alpha} \\ & \,\,\,\,\,\,\,\,\,\,\,\,\,\,\,\,\,\,\,\,\,\,\,\,\,\,\,\,\,\,\,\,\,\,\,\,\,\,\,\,\,\,\,\,\, \quad \text{for $\beta, \beta' =  1,\dots,M$} \\
\end{aligned}
\end{equation}
The first equation above is obtained remembering that $\vec{\rho}$ do not depend on $s$. In Eqs.~\eqref{eq:matrix_direct} and~\eqref{eq:matrix_inverse} the notations
\begin{equation}
\nabla_{\vec{u}} \vec{v} \cdot \nabla_{\vec{u}} \vec{v} = \sum_{\gamma\in\{x,y,z\}}\frac{\p v_\alpha}{\p u_\gamma}\frac{\p v_\beta}{\p u_\gamma} \nonumber
\end{equation}
and
\begin{equation}
\frac{\p \vec{v}}{\p \vec{u}} \cdot \frac{\p \vec{v}}{\p \vec{u}} = \sum_{\gamma\in\{x,y,z\}}\frac{\p v_\gamma}{\p u_\alpha}\frac{\p v_\gamma}{\p u_\beta}, \nonumber
\end{equation}
have been used (both represent $3\times3$ matrices).

\subsection{\textit{Performing the integrals over the constraint variable and momenta}} \label{subsec:AppAIntMom}
The integral over $\pi^s$ in Eq.~\eqref{eq:ConstrainedAverage1} is performed trivially due to the delta function to obtain
\begin{equation}
\begin{aligned}
\Braket{\mathcal{O}} = \frac{1}{\mathcal{Z}'}\int&\dd^{3N}\rho\dd^{3N}\pi^{\rho}\dd^M\sigma \delta^M(\sigma) \mathcal{O}(\vec{\rho}, \sigma, \vec{\pi^{\rho}}, \tilde{\pi}^{\sigma})\\
&\times \delta(\mathcal{H}(\vec{\rho}, \sigma, \vec{\pi^{\rho}}, \tilde{\pi}^{\sigma})-E)\\
\end{aligned}
\end{equation}
The expression above can be usefully simplified by performing the change of variables $\sigma_\alpha \mapsto s_\alpha$, $\vec{\rho}_i \mapsto \vec{R}_i$  to obtain at first
\begin{equation}
\begin{aligned}
\Braket{\mathcal{O}} = \frac{1}{\mathcal{Z}'}\int&\dd^{3N}R\dd^{3N}\pi^{\rho}\dd^Ms |J(\vec{R})| \delta^M(\sigma(\vec{R},s)) \\
&\times\mathcal{O}(R, \sigma(\vec{R},s), \vec{\pi^{\rho}}, \tilde{\pi}^{\sigma})\\
&\times \delta(\mathcal{H}(\vec{R}, \sigma(\vec{R},s), \vec{\pi^{\rho}}, \tilde{\pi}^{\sigma})-E)\\
\end{aligned}
\end{equation}
where $|J(\vec{R})|$ is the Jacobian of the coordinate transformation, which reduces to $\det\left[\frac{\partial \sigma}{\partial s}\right]$. Then, making the dependence on $s$ of the delta explicit, we get
\begin{equation}
\begin{aligned}
\Braket{\mathcal{O}}=\frac{1}{\mathcal{Z}'}\int&\dd^{3N}R\dd^{3N}\pi^{\rho}\dd^Ms |J(\vec{R})| |J(\vec{R})|^{-1} \delta^M(s-\tilde{s}(\vec{R}))\\
&\times O(\vec{R}, s, \vec{\pi^{\rho}}, \tilde{\pi}^{\sigma})\delta(H(\vec{R}, s, \vec{\pi^{\rho}}, \tilde{{\pi}}^{\sigma})-E)\\
\end{aligned}
\end{equation}
where we have $O(\vec{R}, s, \vec{\pi^\rho}, \tilde{\pi}^\sigma) = \mathcal{O}(\vec{R}, \sigma(\vec{R},s), \vec{\pi^{\rho}}, \tilde{\pi}^{\sigma})$.
In this last equality we have used the properties of the delta of a vector function of the integration variable to express the constraint condition directly as a function of the $s$, with $\tilde{s}(\vec{R})$ such that $\sigma(\vec{R},\tilde{s})=0$ (we assume, as commonly done in the Born-Oppenheimer framework that this expression has, for any $\vec{R}$, a single root). Finally, performing the integral over the $s$ variables, and noting that the product of Jacobians in the integrand simplifies, we obtain
\begin{equation}
\begin{aligned}
\Braket{\mathcal{O}} =\frac{1}{\mathcal{Z}'}\int&\dd^{3N} R\dd^{3N}\pi^{\rho}O(\vec{R}, \tilde{s}, \vec{\pi^{\rho}}, \tilde{\pi}^{\sigma}) \\
&\times \delta(H(\vec{R}, \tilde{s}, \vec{\pi^{\rho}}, \tilde{\pi}^{\sigma})-E)\\
\end{aligned}
\end{equation}
with (see next subsection)
\begin{equation}
H(\vec{R}, \tilde{s}, \vec{\pi^{\rho}}, \tilde{\pi}^{\sigma})=\frac{1}{2}{\vec{\pi^{\rho}}}^{\transp}\widetilde{\mathbb{A}}\inv\vec{\pi^{\rho}} + V(\vec{R}, \hat{s})
\end{equation}
which is the result given in the main text. 

\subsection{\textit{Expressions for $\pi^\sigma$ and for the Hamiltonian on the constrained hypersurface}}\label{subsec:AppAHamCon}
The expression for $\pi^\sigma$ on the constrained hypersurface is obtained as follows. From
\begin{eqnarray}
\begin{pmatrix}
\dot{\vec{\rho}} \\
\dot{\sigma}
\end{pmatrix}
 & =
\begin{bmatrix}
\widetilde{\mathbb{\Delta}} & \widetilde{\mathbb{E}} \\
{\widetilde{\mathbb{E}}}\transp & \widetilde{\mathbb{Z}}
\end{bmatrix}
\begin{pmatrix}
\vec{\pi^{\rho}} \\
{\pi}^\sigma
\end{pmatrix}
\end{eqnarray}
where, as in the text, the tildes indicate that quantities are evaluated on the constrained hypersurface. We have
\begin{equation}
\dot{\sigma}=  {\widetilde{\mathbb{E}}}\transp \vec{\pi^{\rho}} +   \widetilde{\mathbb{Z}}\tilde{\pi}^\sigma
\end{equation}
and, since the condition $\dot{\sigma}=0$ must hold on the constrained hypersurface, Eq.~\eqref{eq:ConstrainedMomentum} follows. 

Furthermore, from $\mathbb{M}\mathbb{M}\inv=\mathbb{1}$, we have that  $\widetilde{\mathbb{A}}\widetilde{\mathbb{E}}+\widetilde{\mathbb{B}}\widetilde{\mathbb{Z}}=\mathbb{0}$ and $\widetilde{\mathbb{B}}\transp\widetilde{\mathbb{E}}+\widetilde{\mathbb{\Gamma}}\widetilde{\mathbb{Z}}=\mathbb{1}$. Using these identities, we obtain
\begin{equation}
\begin{aligned}
\widetilde{\mathbb{Z}}\inv&=\widetilde{\mathbb{\Gamma}}-\widetilde{\mathbb{B}}\transp\widetilde{\mathbb{A}}\inv\widetilde{\mathbb{B}} \\
\end{aligned}
\end{equation}    
from which Eq.~\eqref{eq:ConstrainedMomentumNew} is obtained by substituting the expression above in Eq.~\eqref{eq:ConstrainedMomentum}.

The Hamiltonian
\begin{equation}
\mathcal{H}(\upsilon, \pi^{\upsilon}) =  \frac{1}{2}\pi^{\upsilon \transp} \mathbb{M}^{-1}(\upsilon)\pi^{\upsilon} + \mathcal{V}(\upsilon) \nonumber
\end{equation}
on the hypersurface $\sigma=0$, $\pi^{\sigma}=\tilde{\pi}^{\sigma}$ can be written in the form presented in the text via the following steps.
First, we use the block representation of the inverse mass matrix to write
\begin{equation}
\begin{aligned}
& \mathcal{H}(\vec{\rho}, \sigma = 0, \vec{\pi^{\rho}}, \pi^{\sigma} = \tilde{\pi}^\sigma) \\
 & = \frac{1}{2}(\vec{\pi^{\rho}}, \tilde{\pi}^\sigma)
\begin{bmatrix}
\widetilde{\mathbb{\Delta}} & \widetilde{\mathbb{E}} \\
{\widetilde{\mathbb{E}}}\transp & \widetilde{\mathbb{Z}}
\end{bmatrix}
\begin{pmatrix}
\vec{\pi^{\rho}} \\
\tilde{\pi}^\sigma
\end{pmatrix}
+ \mathcal{V}(\vec{\rho}, \sigma=0) \\
\end{aligned}
\end{equation}
We then observe that 
the block expression of the product $\mathbb{M}\mathbb{M}\inv=\mathbb{1}$ imposes
$\widetilde{\mathbb{A}}\widetilde{\mathbb{\Delta}}+\widetilde{\mathbb{B}}\widetilde{\mathbb{E}}\transp=\mathbb{1}$ and $\widetilde{\mathbb{A}}\widetilde{\mathbb{E}}+\widetilde{\mathbb{B}}\widetilde{\mathbb{Z}}=\mathbb{0}$. These two relationships, in turn, imply $\widetilde{\mathbb{\Delta}} - \widetilde{\mathbb{E}} {\widetilde{\mathbb{Z}}}\inv{\widetilde{\mathbb{E}}}\transp=\widetilde{\mathbb{A}}\inv$ so that
\begin{equation}
\begin{aligned}
& \mathcal{H}(\vec{\rho}, \sigma = 0, \vec{\pi^{\rho}}, \pi^{\sigma} = \tilde{\pi}^\sigma)  \\
& =  \frac{1}{2}
[{\vec{\pi^{\rho}}}^{\transp}(\widetilde{\mathbb{\Delta}} - \widetilde{\mathbb{E}}{\widetilde{\mathbb{Z}}}\inv{\widetilde{\mathbb{E}}}\transp)\vec{\pi^{\rho}}] + \mathcal{V}(\vec{\rho}, \sigma=0) \\
& = \frac{1}{2}
{\vec{\pi^{\rho}}}^{\transp}\widetilde{\mathbb{A}}\inv\vec{\pi^{\rho}} + \mathcal{V}(\vec{\rho}, \sigma=0) 
\end{aligned} 
\end{equation}

\section{Interaction potentials used in simulations}
\label{app:ShellModel}

Let us indicate with $\vec{R}=\{\vec{R}_1,\dots,\vec{R}_N\}$ and $\vec{S}=\{\vec{s}_1,\dots,\vec{s}_N\}$, where $N$ is the number of ions. The Shell Model (SM) potential adopted in our calculations is of the form
\begin{equation}
\begin{aligned}
&V^{\text{SM}}(\vec{R},\vec{S}) = \sum_{i=1}^N \frac{1}{2}k_i(\vec{R}_i - \vec{s}_i)^2 \\
&+ \sum_{i\ne j} \biggl\{\frac{q_i q_j}{|\vec{s}_i - \vec{s}_j|} + \frac{q_iQ_j}{|\vec{s}_i - \vec{R}_j|} + \frac{Q_iq_j}{|\vec{R}_i - \vec{s}_j|} + \frac{Q_iQ_j}{|\vec{R}_i - \vec{R}_j|} \\
&+ A_{ij}\exp\biggr[-\frac{|\vec{s}_i - \vec{s}_j|}{\lambda}\biggl] - \frac{C_{ij}}{|\vec{R}_i - \vec{R}_j|^6} - \frac{D_{ij}}{|\vec{R}_i - \vec{R}_j|^8}\biggr\}
\end{aligned}
\end{equation}
where $Q_i$ and $q_i$ represent the charge assigned to the core and the shell of the ion $i$, respectively\footnote{The parameters $Q_i$ and $q_i$ are subject to the condition $Z_i = Q_i + q_i$, where $Z_i$ is the ionic charge.} and $k_i$, $A_{ij}$, $\lambda = 0.317\um{\angstrom}$, $C_{ij}$ and $D_{ij}$ are force-field parameters dependent on the particular species considered. The values chosen for these parameters are reported in Table~\ref{tab:coefficients_inter} and Table~\ref{tab:coefficients_single}. 
\begin{table}
\caption{Interatomic potential parameters for the Shell Model and for the Rigid Ion Model from Ref.~\cite{jacucci:1976}.}
\label{tab:coefficients_inter}       
\begin{tabular}{llll}
\hline\noalign{\smallskip}
species & $A_{ij}$ $(\um{\hartree})$ & $C_{ij}$ $(\um{\hartree}\um{\angstrom^{-6}})$ & $D_{ij}$ $(\um{\hartree}\um{\angstrom^{-8}})$  \\
\noalign{\smallskip}\hline\noalign{\smallskip}
Na Na & $15.57$ & $3.85\cdot10^{-2}$  & $1.83\cdot10^{-2}$ \\
Na Cl & $46.10$ & $2.57\cdot10^{-1}$ & $3.19\cdot10^{-1}$ \\
Cl Cl & $128.0$ & $2.66$ & $5.34$\\
\noalign{\smallskip}\hline
\end{tabular}
\end{table}
\begin{table}
\caption{Single species parameter for the Shell Model from Refs.~\cite{jacucci:1976,ishii:2015}.}
\label{tab:coefficients_single}       
\begin{tabular}{llll}
\hline\noalign{\smallskip}
species & $k_i$ $(\um{\hartree}\um{\angstrom^{-2}})$ & $Q_i$ $(\um{\elec})$ & $q_i$ $(\um{\elec})$  \\
\noalign{\smallskip}\hline\noalign{\smallskip}
Na & $35.1$ & $+3.17$ & $-2.17$ \\
Cl & $1.58$ & $+2.17$ & $-3.17$ \\
\noalign{\smallskip}\hline
\end{tabular}
\end{table}
The analytical form of the potential is the same as the one used in Ref.~\cite{jacucci:1976} but, differently from that work that considered a mixed picture with rigid ion Na$^+$ and polarizable Cl$^-$, we consider a model in which both species are polarizable. Consequently, the parameters in the potential associated to the shell variables paired with Sodium are nonzero. The values used for these parameters are based on a more sophisticated polarizable potential proposed in~\cite{ishii:2015} (in particular, we used that work to define the elastic constant $k$) and the Sodium core and shell charges were assigned by mirroring (with opposite signs) those used for Chlorine. The changes caused by this reparametrization of the shell model potential are minor as demonstrated in Section~\ref{sec:Results}. 

The Rigid Ion Model (RIM) used for the simulations illustrated in Figure~\ref{fig:PolNoPolDiff} and in the subsequent discussion is defined by
\begin{equation}
\begin{aligned}
V^{\text{RIM}}(\vec{R}) &= \sum_{i\ne j} \biggl\{\frac{Z_iZ_j}{|\vec{R}_i - \vec{R}_j|} + A_{ij}\exp\biggr[-\frac{|\vec{R}_i - \vec{R}_j|}{\lambda}\biggl] + \\
&- \frac{C_{ij}}{|\vec{R}_i - \vec{R}_j|^6} - \frac{D_{ij}}{|\vec{R}_i - \vec{R}_j|^8}\biggr\}
\end{aligned}
\end{equation}
where the parameters $A_{ij}$, $C_{ij}$, $D_{ij}$ and $\lambda$ are the same as the shell model and the ionic charges $Z_i = Q_i + q_i$ are used in place of the ones summarized in Table~\ref{tab:coefficients_single}.

\end{document}